\documentclass[aps,reprint,groupaddress,superscriptaddress,amssymb,prx]{revtex4-2}
\usepackage{physics}
\usepackage{enumitem}
\usepackage{stackengine}
\usepackage{xcolor}
\usepackage{mathtools}
\usepackage{floatrow}
\usepackage{subfig}
\usepackage{graphicx}
\usepackage{dcolumn}
\usepackage{bm}
\usepackage{hyperref}
\usepackage[mathlines]{lineno}
\usepackage{prettyref}
\usepackage{soul}
\usepackage{placeins}
\usepackage{xurl}
\usepackage{comment}
\definecolor{linkcolor}{rgb}{0,0,0.65}
\definecolor{linescolor}{rgb}{0.65,0.1,0.1}
\definecolor{cool}{RGB}{49,54,149}
\definecolor{hot}{RGB}{165,0,38}
\hypersetup{ colorlinks=true, pdfstartview=FitV,
             linkcolor=linescolor, citecolor=linescolor,
             urlcolor= linkcolor } 
\newcommand{\quotes}[1]{``#1''}
\newrefformat{eq}{Eq.~(\ref{#1})}
\newrefformat{expr}{(\ref{#1})}
\newrefformat{fig}{Figure \ref{#1}}
\newrefformat{tab}{Table \ref{#1}}
\newrefformat{app}{\ref{#1}}

\begin{document}

\preprint{APS/123-QED}

\title{Finite size scaling of survival statistics in metapopulation models}

\author{Alice Doimo}
\thanks{Corresponding author: alice.doimo@phd.unipd.it}
\affiliation{Laboratory of Interdisciplinary Physics, Department of Physics and Astronomy \quotes{G. Galilei}, University of Padova, Via Marzolo 8, 35131 Padova, Italy}
\affiliation{National Biodiversity Future Center, Piazza Marina 61, 90133, Palermo, Italy}
\author{Giorgio Nicoletti}
\affiliation{Quantitative Life Sciences section, The Abdus Salam International Center for \\ Theoretical Physics (ICTP), Trieste, Italy}
\affiliation{ECHO Laboratory, École Polytechnique Fédérale de Lausanne, Lausanne, Switzerland}
\author{Davide Bernardi}
\affiliation{Laboratory of Interdisciplinary Physics, Department of Physics and Astronomy \quotes{G. Galilei}, University of Padova, Via Marzolo 8, 35131 Padova, Italy}
\affiliation{National Biodiversity Future Center, Piazza Marina 61, 90133, Palermo, Italy}
\author{Prajwal Padmanabha}
\affiliation{Department of Fundamental Microbiology, University of Lausanne, Switzerland}

\begin{abstract}
    Spatial metapopulation models are fundamental to theoretical ecology, enabling to study how landscape structure influences global species dynamics. Traditional models, including recent generalizations, 
     often rely on the deterministic limit of stochastic processes, assuming large population sizes. 
    However, stochasticity - arising from dispersal events and population fluctuations - profoundly shapes ecological dynamics.
    In this work, we extend the classical metapopulation framework to account for finite populations, examining the impact of stochasticity on species persistence and dynamics. 
    Specifically, we analyze how the limited capacity of local habitats influences survival, deriving analytical expressions for the finite-size scaling of the survival probability near the critical transition between survival and extinction. 
    Crucially, we demonstrate that the deterministic metapopulation capacity plays a fundamental role in the statistics of survival probability and extinction time moments. 
    These results provide a robust foundation for integrating demographic stochasticity into classical metapopulation models and their extensions.
\end{abstract}

\maketitle

\section{\label{sec:intro}introduction}
\noindent Understanding the dynamics of populations distributed across fragmented habitats is a longstanding challenge in ecology and conservation biology.
Metapopulation theory provides a foundational framework for examining how local extinctions and landscape-mediated colonization processes shape the collective dynamics of spatially structured populations over time
\cite{leibold2004metacommunity, Hanski1, Hanski1991SinglespeciesMD, spoms_and_ifm, metapop, hanski1998connecting, wang2015dispersal, loreau2003meta, loreau2003biodiversity, marleau2014meta}. The concept of metapopulation, first introduced by Levins \cite{levins_def}, describes a \quotes{population of populations}, i.e., a set of distinguished subpopulations that are spatially separated, but interconnected by the ongoing exchange of individuals. This exchange occurs across a spatial network of habitat patches, varying in quality, connectivity, and area \cite{Hanski1}. Both field and theoretical studies have highlighted the validity of this approach, showing that landscape topology significantly influences the flow of individuals between habitats, ultimately determining the survival or extinction of the metapopulation \cite{exp1,exp2,exp3,exp4,exp5}.

In particular, to capture the effects of spatial structure and habitat fragmentation on species persistence, Hanski and Ovaskainen introduced a fundamental measure known as metapopulation capacity \cite{metapoplandscape, Hanski2, HO_extinctionthr}, which determines the survival of a focal species within a given landscape. This measure is defined as the maximum eigenvalue of an appropriate landscape matrix, and allows for a direct comparison between theoretical models and real-world networks of habitat fragments \cite{ho2004}.
Although the application of network theory to spatial ecology \cite{marquet05, montoya2006ecological, fortin2021network} has advanced the study of complex dispersal network structures \cite{metapop, white2010population, holland2008strong, gilarranz2012spatial, padmanabha2024spatially, grilli2015metapopulation}, metapopulation capacity has traditionally been defined within deterministic models. Crucially, such models overlook fluctuations arising from demographic and environmental stochasticity \cite{bonsall2004demographic, mangel1993dynamics, lai2011effects}.
Early contributions by Lande \cite{Lande} and Hanski \cite{practicalmodelMETAPOP} pioneered the integration of stochasticity into metapopulation dynamics, emphasizing the increased risk of extinction posed by random fluctuations at low population levels. Migration through spatially connected patches, by contrast, provides insurance against simultaneous global extinction, underscoring the fundamental role of dispersal network topology. Classical approaches to studying these dynamics, such as Stochastic Patch Occupancy Model \cite{day1995stochastic,moilanen2004spomsim}, the Incidence Function Model \cite{practicalmodelMETAPOP} and their extensions \cite{verboom1991linking,etienne2001,ross2006,sutherland2014demographic}, have enabled the computation of extinction times and quasi-stationary occupancy distributions. However, these approaches predominantly rely on purely numerical methods due to the complexity of stochastic processes, spatial heterogeneity, and variability in the parameter space \cite{tao2024landscape}.
As a consequence, in metapopulation dynamics, qualitative results are often obtained from deterministic approximations of stochastic processes, which lead to theoretically tractable models and provide broad qualitative insights \cite{spoms_and_ifm}.

In this work, we rigorously extend the traditional deterministic metapopulation framework by explicitly incorporating the stochasticity of individual-level processes to investigate analytically their potential impact on metapopulation dynamics and persistence. To incorporate stochasticity from the ground up, we focus on the role of finite carrying capacity, i.e., the maximum size of the subpopulations that compose the metapopulation. This approach offers a precise way to control fluctuations, allowing us to disentangle the effects of local demographic stochasticity and dispersal network topology.
Specifically, we examine whether critical thresholds in carrying capacity exist beyond which stochastic fluctuations become catastrophic, or conversely, whether there are local system sizes beyond which population persistence is significantly enhanced. To address this question, we build upon our previous work, which extends the deterministic spatial metapopulation model developed by Hanski and Ovaskainen to arbitrary landscape structures \cite{metapop}. We adopt a bottom-up approach, starting from a microscopic model describing colonization, extinction, and dispersal of a focal species at the individual level.

To analytically study the stochastic dynamics of the system, we perform a Van Kampen inverse system-size expansion of the corresponding master equation \cite{VanKampen}, selecting the inverse of the carrying capacity of the patches - i.e., the maximum local population size supported by each patch - as the expansion parameter. We then introduce a series of simplifying hypotheses to derive analytical insights from the resulting Fokker-Planck equation. In particular, we assume translational invariance within a complete network and marginalize to capture the effective behavior of representative single-site variables.
By imposing a separation of timescales \cite{nicoletti2024information} between dispersal and local processes, we derive an effective one-dimensional quasi-stationary (QSS) Langevin equation that accurately captures the overall metapopulation dynamics.
In the deterministic limit, we recover the expected absorbing phase transition \cite{grinstein1997statistical, munoz1997survival} between extinction and survival regimes, as determined by the system’s metapopulation capacity. Stochasticity, however, renders the survival regime metastable - a key result especially relevant for systems with small carrying capacity. Moreover, the effective QSS equation enables us to analytically study the general scaling behavior of the survival probability in the vicinity of the extinction transition. We demonstrate that near this transition point, the survival probability for different carrying capacities exhibits a universal behavior consistent with finite-size scaling \cite{finitesizescaling1, finitesizescaling2}. Furthermore, we analytically derive the scaling of the first and second moments of extinction time with varying carrying capacity, showing that they similarly collapse to a universal form.

Through this analysis, we bridge results from the deterministic metapopulation framework with its stochastic counterpart. We demonstrate that the metapopulation capacity - a deterministic measure - still plays a crucial role in the statistics of survival probability and extinction time moments.

\section{\label{sec:det} model}
\noindent 
We introduce a model that describes the microscopic dynamics of a single species within a metapopulation. Individuals undergo processes such as birth, death, reproduction, and dispersal across a network of habitat patches. This model was proposed in \cite{metapop}, where the connection between its deterministic limit and the Hanski and Ovaskainen model \cite{Hanski1,Hanski2} was demonstrated. In this section, we first provide a detailed description of the microscopic processes governing the dynamics and then briefly recall the derivation of the macroscopic deterministic equations, as outlined in \cite{metapop}. This is instrumental in the analysis of the full stochastic model, which we analyze in the following section.
\subsection{Microscopic ecological dynamics\label{sec:micro}}
\noindent 
We consider a dispersal network of $N$ interconnected habitat patches, corresponding to the nodes of the dispersal network, inhabited by a given focal species. We separately model the behavior of two kinds of individuals: settled individuals (denoted by $S$), which reside on the nodes of the network, and explorers ($X$), which diffuse through the edges of the network.
We denote the vector comprising all local abundances of individuals in each node by
$\vec{n} = (n_{X_1},... n_{X_N};n_{S_1},... n_{S_N})$, where $n_{X_i}$ and $n_{S_i}$ are the numbers of explorers and settled individuals in patch $i$, respectively. 
By assuming that each patch has a maximum carrying capacity of $M$ settled individuals - i.e., $ n_{S_i} + n_{\varnothing_i} = M \quad \forall i$, where $\varnothing_i$ denotes an empty site in patch $i$ - we define the following microscopic reactions:
\vspace{-2em}
\begin{equation}
\begin{gathered}
     S_i \xrightarrow[]{c_{ij}} S_i + X_j \qquad
     S_i \xrightarrow[]{e_i} \varnothing_i \\
     X_i+\varnothing_i \xrightarrow[]{\lambda/M} S_i \qquad
     X_i+S_i \xrightarrow[]{\lambda/M} S_i \qquad
     X_i \xrightarrow[]{\mathcal{D}_{ij}} X_j 
\end{gathered}
\label{eq:reactions}
\end{equation}
where the indices $i,j=1,...N$ label the network patches.
The rate $e_i$ accounts for the death of a settled individual in patch $i$, while $c_{ij}$ represents the rate at which an explorer in patch $j$ is produced by a settled individual in a neighboring node $i$.
An explorer in $i$ can move to a connected patch $j$ at rate $\mathcal{D}_{ij}$, and it may attempt to colonize a randomly chosen site in its current patch at a rate $\lambda/M$. In this process, the explorer does not necessarily target only available sites. Rather, if the chosen site is empty, the explorer settles, whereas if the site is already occupied, the explorer dies. This formulation ensures that the total rate of explorer removal reflects both successful colonization and death due to an occupied site, while the rescaling of $\lambda$ with $M$ guarantees that such rate is independent of $M$.
Diffusion rates are set in the form $\mathcal{D}_{ij}=\mathcal{D} \,\mathcal{A}_{ij}$ with $\mathcal{A}_{ij}$ the adjacency matrix of the dispersal network, denoting a possibly weighted connection between node $j$ and node $i$. Similarly, the colonization rate $c_{ij}$ is given by $c_{ij} = c_i \,h(\mathcal{D}/\lambda)\, \mathcal{A}_{ij}$, where $h(f)$ is a function that encodes feasibility of exploration and $f=\mathcal{D}/\lambda$ controls exploration efficiency. A Monod function $h(f) = \frac{\xi_0}{1+1/f}$ is typically chosen to model the saturation effect of exploration feasibility, where $\xi_0$ represents the maximal explorability.

\subsection{Deterministic equations}
\noindent The deterministic equations resulting from the microscopic model defined in \prettyref{eq:reactions} are given by:
\begin{gather}
\label{eq:det}
        \dot{\rho}_i =  - e_i\,\rho_i +\lambda \,( 1- \rho_i ) \,x_i \\
        \dot{x}_i = - \lambda  x_i + h(f) \sum_{j=1}^N \mathcal{A}_{ji} c_j \rho_j 
        + \mathcal{D}\sum_{j=1}^N ( \mathcal{A}_{ji} x_j - \mathcal{A}_{ij} x_i) \nonumber
\end{gather}
where $\rho_{i},\, x_{i}$ are the average population densities, $\langle  n_{S_i}\rangle/M$ and $\langle  n_{X_i}\rangle/M$, respectively.
In the limit in which the explorers' dynamics is much faster than the dynamics of the settled population, one obtains an effective model describing the evolution of the settled population:
\begin{equation}\label{eq:QSSdet}
    \dot{\rho}_i =  -e_i \rho_i +(1-\rho_i)\sum_{j=1}^N c_j K_{ji} \rho_j
\end{equation}
where the kernel matrix $K$ encodes the dependence of the population density on the topology of the dispersal network. We have that
\begin{equation}
    K_{ji}= \sum_{k=1}^N\, C_{jk}F^{-1}_{ik}
\end{equation}
where we introduced the matrix $F= \mathbb{I} + f L^T$ that depends on the out-degree Laplacian of the network, $L_{ij}= \delta_{ij}\sum_{k=1} ^N A_{ik}-A_{ij}$. The kernel matrix, $K$, introduces an effective coupling between settled populations residing in different patches. In particular, in \cite{metapop} it was shown that
\begin{equation}
    \label{eqn:kernel}
    K_{ji} = h(f)\sum_{l=1}^N A_{jl} \sum_{k = 1}^N \frac{V_{ik} (V^{-1})_{kl}}{1 + f \omega_k}
\end{equation}
where $\omega_k$ is the $k$-th eigenvalue of the transpose out-degree Laplacian of the dispersal network, and $V_{ij} = v_i^{(j)}$ with $\vb{v}^{(k)}$ the corresponding right eigenvector. Furthermore, from the deterministic model in \prettyref{eq:QSSdet}, one can show that if $e_i = e$ and $c_i = c$ for all patches, the survival of the species is uniquely determined by the maximum eigenvalue of the dispersal kernel, i.e., the metapopulation capacity $\lambda_M$ proposed by Hanski and Ovaskainen \cite{metapoplandscape}. If $\lambda_M > e/c$, the species survives; otherwise, it goes extinct. A similar condition has been shown to hold when $e_i$ explicitly depends on the patch index \cite{padmanabha2024spatially}. 


\begin{figure*}[t]
    \centering
    \includegraphics[width=0.9\linewidth]{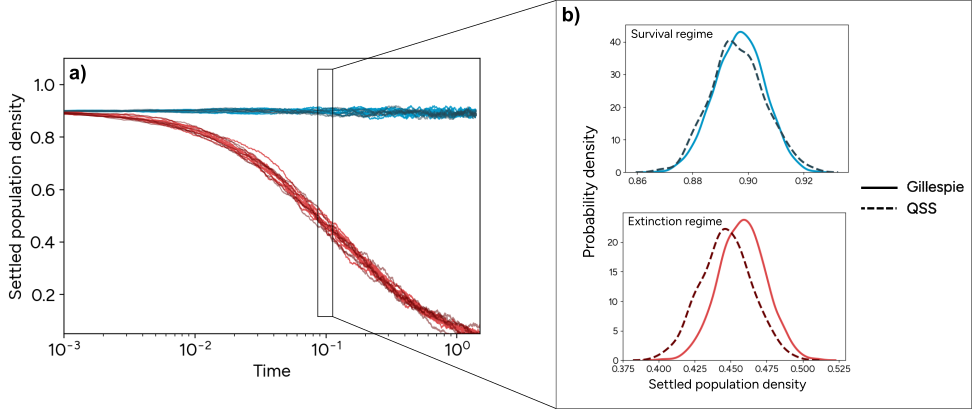} 
    \caption{
    \raggedright
     Comparison between \textbf{a)} 5 realizations of the trajectories resulting from the mean-field quasi-stationary SDE \prettyref{eq:MFeq} (integrated with the  Euler-Maruyama method) and the Gillespie simulation reproducing the exact dynamics of the microscopic model \prettyref{eq:reactions}. The parameters of the model are $M=1000$, $N=20$, $f=1$ which imply an effective metapopulation capacity of $\lambda_M \sim 9.5$.
    Survival regime ($e/c<\lambda_M$): $e=1$, $c=1.5$, 
    extinction regime ($e/c>\lambda_M$): $e=10$, $c = 1$.
    }
    \label{fig:gillespie}
\end{figure*}

\section{\label{sec:stoch}stochastic model}


\noindent In deriving the deterministic dynamics given by \prettyref{eq:det}, we have implicitly assumed a large carrying capacity, i.e., $M\to\infty$.
However, to investigate the effect of stochasticity on population persistence, we need to extend the deterministic model to account for finite carrying capacities, which introduce demographic fluctuations.
For a general model defined by a set of microscopic processes, the complete stochastic dynamics is governed by the master equation \cite{Gardiner}
\begin{align}\label{eq:ME}
    \partial_t \mathcal{P}(\vec{n},t) = \sum _{\vec{n}' \neq \vec{n}} \biggl[ &
    W(\vec{n}|\vec{n}') \mathcal{P}(\vec{n}',t) + \nonumber \\
    & - W(\vec{n}'|\vec{n}) \mathcal{P}(\vec{n},t) \biggr]
\end{align}
where  $\mathcal{P}(\vec{n},t)$ is the probability that the system is in state $\vec{n}$ at time $t$, and $W(\vec{n}'|\vec{n})$ are the rates at which the system transitions from  state $\vec{n}$ to state $\vec{n}'$. The transition rates for our model can be straightforwardly derived from the reactions \prettyref{eq:reactions}, leading to:
\begin{align}\label{eq:Wrates}
    & W(n_{S_i}-1;n_{X_i}|n_{S_i};n_{X_i}) = e_i \, n_{S_i} \nonumber \\
    & W(n_{S_i}+1;n_{X_i}-1|n_{S_i};n_{X_i}) = \frac{\lambda}{M}  \, n_{X_i} (M-n_{S_i}) \nonumber \\
    & W(n_{S_i};n_{X_i}-1|n_{S_i};n_{X_i}) = \frac{\lambda}{M} \, n_{X_i} n_{S_i} \\
    & W(n_{S_i}, n_{S_j}; n_{X_i}+1,n_{X_j} |n_{S_i}, n_{S_j}; n_{X_i}, n_{X_j}) = c_{ji}  n_{S_j} \nonumber \\
    & W(n_{S_i}, n_{S_j}; n_{X_i}+1, n_{X_j}-1|n_{S_i}, n_{S_j}; n_{X_i}, n_{X_j}) = \mathcal{D}_{ji}  n_{X_j} \nonumber
\end{align}
where, for the sake of brevity, we explicitly write only the number of individuals at the nodes involved in the transition.

\subsection{\label{sec:FP}Fokker-Planck equation}
\noindent Although no closed-form solution for the master equation of our model exists, the corresponding stochastic dynamics can be simulated exactly using the Gillespie algorithm \cite{gillespie}. However, we can obtain analytical insight into the effects of stochasticity by means of a Van Kampen system size expansion \cite{vankampenexp, VanKampen}. We take $1/M$ as the natural expansion parameter, so that in the $M \to \infty$ limit we recover the deterministic equations.
To perform this expansion, it is convenient to rewrite the master equation in terms of step operators, defined by their action on a generic function $f(\vec{n})$ as:
\begin{equation}
\xi^{\pm1}_{A_i} 
f(n_{A_1},... n_{A_i},... n_{A_N}) = 
f(n_{A_1},... n_{A_i}\pm1,... n_{A_N}) 
\end{equation} 
(see Appendix \prettyref{app:step_operators} for details).
In terms of step operators, the master equation becomes:
\begin{widetext}
\begin{equation}
\label{eq:master}
\begin{aligned}
    \frac{\partial}{ \partial t}\mathcal{P}(\vec{n}, t)  = &
    \sum_{i=1}^N \Big \{ \Big[
    \left( \xi_{S_i}^{+1} - 1\right)
    W(n_{S_i}-1;n_{X_i}|n_{S_i};n_{X_i}) 
     +
    \left( \xi_{S_i}^{-1}\xi_{X_i}^{+1} - 1\right)
    W(n_{S_i}+1;n_{X_i}-1|n_{S_i};n_{X_i})
    \,   \\&
    +\left(\xi_{X_i}^{+1} - 1\right)
    W(n_{S_i};n_{X_i}-1|n_{S_i};n_{X_i}) \Big ]
    \mathcal{P}(n_{S_i};n_{X_i}) \Big \}
    \, \\& +
    \sum_{i=1}^N  \sum_{j=1}^N
    \Big \{
    \Big [ 
    \left(\xi_{X_i}^{-1} - 1\right)
    W(n_{S_i}, n_{S_j}; n_{X_i}+1, n_{X_j}|n_{S_i}, n_{S_j}; n_{X_i}, n_{X_j}) 
    \,   \\& +
    \left(\xi_{X_i}^{-1}\xi_{X_j}^{+1} - 1\right)
    W( n_{S_i}, n_{S_j}; n_{X_i}+1, n_{X_j}-1|n_{S_i}, n_{S_j}; n_{X_i}, n_{X_j}) 
    \Big ] 
    \mathcal{P}( n_{S_i}, n_{S_j}; n_{X_i}, n_{X_j})
   \Big \}  \;.
\end{aligned} 
\end{equation}
Note that the step operators act on the entire product \( W \mathcal{P} \) (i.e., both on \( W \) and \( \mathcal{P} \)), but we have grouped \( \mathcal{P} \) to keep the expression more compact.
If we define the population densities with respect to the system size parameter $M$ as 
$ \rho_i = n_{S_i}/M   , \, x_i = n_{X_i}/M $, 
the expansion of the master equation up to second order in powers of $1/M$, for fixed densities, yields the Fokker-Planck (FP) equation:
\begin{align}\label{eq:fullFPeq}
\partial_t \mathcal{P}(\vec{\rho},\vec{x},  t)  = &
    - \sum_{i=1}^N \partial_{\rho_i} 
    \left[ A^{\rho}_i(\vec{\rho},\vec{x})\mathcal{P}(\vec{\rho},\vec{x}, t)
    \right] 
    - \sum_{i=1}^N \partial_{x_i} 
    \left[ A^{x}_i(\vec{\rho},\vec{x})\mathcal{P}(\vec{\rho},\vec{x}, t)
    \right] 
    + \frac{1}{2} 
     \sum_{i,j=1}^N \partial_{\rho_i} \partial_{\rho_j} \left[ \mathbb{D}^{\rho\rho}_{i,j}(\vec{\rho},\vec{x}) \mathcal{P}(\vec{\rho},\vec{x},  t) \right] 
    +
   \nonumber \\ &
    + \frac{1}{2} 
     \sum_{i,j=1}^N \partial_{x_i} \partial_{x_j} \left[ \mathbb{D}^{xx}_{i,j}(\vec{\rho},\vec{x})\mathcal{P}(\vec{\rho},\vec{x},  t) \right]
     +\sum_{i,j=1}^N \partial_{\rho_i} \partial_{x_j} \left[ \mathbb{D}^{\rho x}_{i,j}(\vec{\rho},\vec{x}) \mathcal{P}(\vec{\rho},\vec{x},  t) \right] \; .
\end{align}
\end{widetext}
Here, the subscripts $i,j,k$ label the network patches while the superscripts $\rho, x$ refer to the settled and explorer population, respectively.
The drift vector's components correspond to the deterministic model:
\begin{align}\label{eq:drift}
    & A^{\rho}_i = 
     \lambda \, x_i (1-\rho_i) -  e_i \rho_i 
    \nonumber  \\
    & A^{x}_i = 
    \sum_j 
    \left(
     \mathcal{D}_{ji} x_j - \mathcal{D}_{ij} x_i  
    + c_{ji} \rho_j \right) - \lambda \,x_i
\end{align}
and the Fokker-Planck diffusion matrix has the following block structure
\begin{align}\label{eq:diffcomponents}
  &\mathbb{D} = 
  \begin{pmatrix}
      \mathbb{D}^{\rho\rho} 
      & \mathbb{D}^{\rho x} \\
      \mathbb{D}^{ x \rho} 
      &  \mathbb{D}^{xx}
 \end{pmatrix}
 \end{align}
with blocks given by
\begin{align}
    \mathbb{D}^{\rho\rho}_{ij} = & \frac{1}{M}\left[ e_i \rho_i + \lambda  x_i (1 - \rho_i) \right]\delta _{ij} 
    \nonumber \\
    \mathbb{D}^{\rho x}_{ij} = & \mathbb{D}^{x \rho}_{ij} = -\frac{1}{M} \lambda x_i (1-\rho _i)  \delta _{ij} 
    \nonumber \\
    \mathbb{D}^{xx}_{ij} = & \frac{1}{M}\left[\sum_k (\rho _k c_{ki}+x_i \mathcal{D}_{ik}+x_k \mathcal{D}_{ki})+\lambda  x_i\right] \delta_{ij} + 
    \nonumber \\
    & -\frac{1}{M}(x_i \mathcal{D}_{ij}+x_j \mathcal{D}_{ji}) (1- \delta _{ij}) \; .
\end{align}
where $\delta_{ij}$ is the Kronecker delta. We highlight that all entries in the Fokker-Planck diffusion matrix are proportional to the inverse of the carrying capacity, explicitly showing that the leading order of the Van Kampen expansion yields the deterministic dynamics.

\begin{figure*}[t]
    \centering
    \subfloat[]{
    \includegraphics[width=0.87\linewidth]{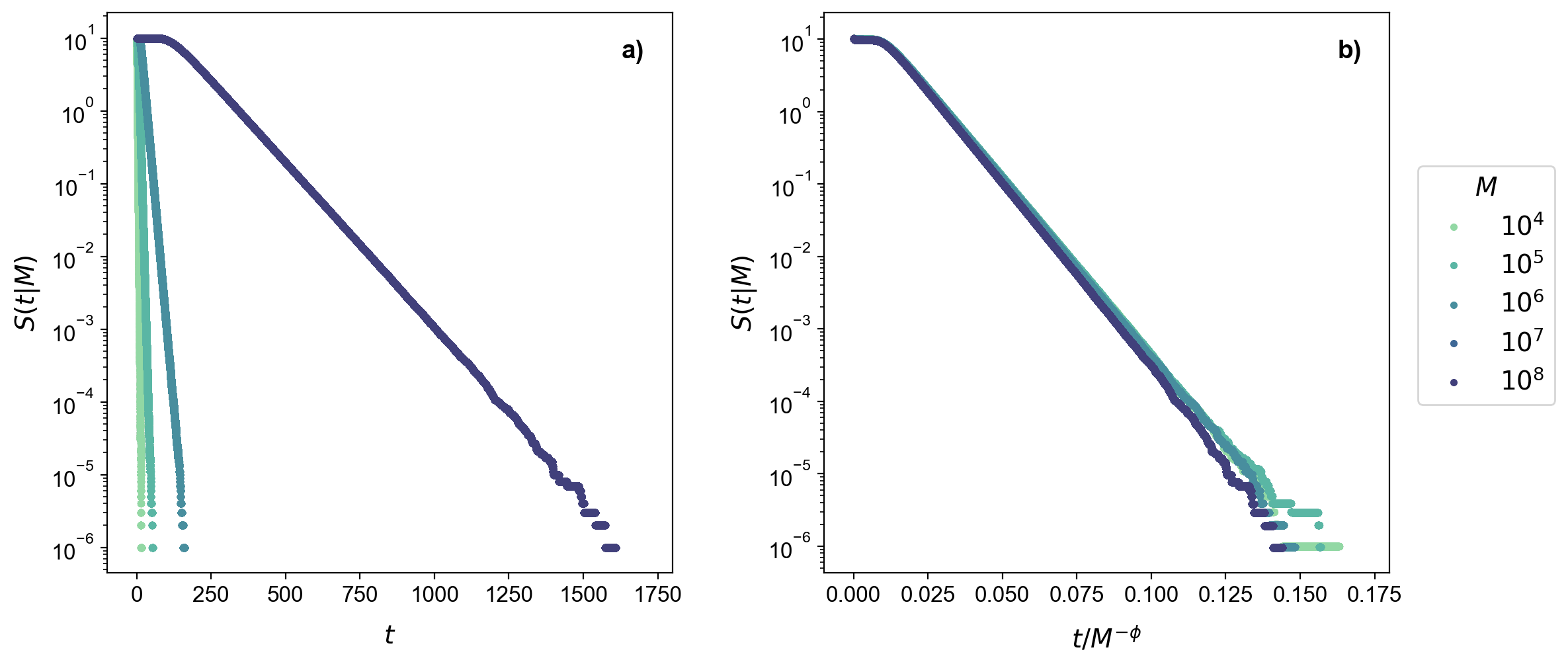} 
    }
    \caption{
    \raggedright
     Survival probability of the quasi-stationary SDE as a function of time $t$ and system size $M$: \textbf{(a)} not rescaled. \textbf{(b)} rescaled according to \prettyref{eq:hyp}, with the numerical estimates of the scaling exponents given in \prettyref{eq:fittedexp}. The system is set at criticality and the initial condition is fixed at $\rho=\frac{1}{2}$. The microscopic parameters of the model are 
    \footnotesize
    $N = 201$, $\lambda = 100$, $D = 100$, $e/c=\lambda_M=100$.
    \small
    Averages are obtained over $10^7$ numerical realizations of the dynamics in \prettyref{eq:MFeq}, using Euler-Maruyama algorithm.}
    \label{fig:scaling1}
\end{figure*}

Although \prettyref{eq:fullFPeq} provides an analytical description of our stochastic model in the limit of large system size $M$, a general analytical treatment remains cumbersome for large networks, as it would require integrating $2N$ variables. Additionally, the explorers' noise strength, $\mathbb{D}^{xx}$, has non-zero off-diagonal entries, coupling their dynamics across patches. Hence, we resort to a further simplification of \prettyref{eq:fullFPeq} to obtain a more direct understanding of the system's behavior.

\subsection{Effective SDE}\label{sec:QSS}
\noindent 
To simplify the dynamics and reduce the number of variables, we assume a fully connected dispersal network $\mathcal{A}_{ij} = \mathcal{A}(1-\delta_{ij})$, and homogeneous habitat patches $c_i = c$, $e_i = e \,\, \forall \, i$, implying that the system is invariant under translations. $\mathcal{A}$ is set to $1/N$ to ensure a well-defined limit for $N\rightarrow \infty$. 
\\
To gain analytical insight, we focus on the marginalized probability distribution, by integrating out all variables except those associated with a single site, $(\rho_1,x_1) \equiv (\rho, x)$:
\begin{align}\label{eq:margP}
    \int \prod_{i \geq 2} \left[d\rho_i dx_i \right] \, \mathcal{P}(\vec{\rho}, \vec{x}, t) = P(\rho, x, t) \,.
\end{align}
Motivated by our assumption of a fully connected dispersal network, we adopt the following approximation for each $\rho_j$, $x_j$:
\begin{align}\label{eq:margappr}
    & \int  \prod_{i \geq 2} \left[d\rho_i dx_i \right] \, \mathcal{P}(\vec{\rho}, \vec{x}, t) \, x_j \approx  P(\rho, x, t) \, x
    \nonumber \\
    & \int \prod_{i \geq 2} \left[d\rho_i dx_i \right] \, \mathcal{P}(\vec{\rho}, \vec{x}, t) \, \rho_j \approx  P(\rho, x, t) \, \rho
\end{align}
Finally, we take the limit $N\rightarrow \infty$, obtaining the simplified equation:
\begin{equation}\label{eq:MFFP}
\begin{split}
\frac{\partial}{\partial t} \,  P(\rho,x,t) \,  
 = \,
- \, \partial_{\rho}\left \{ \left[\,\lambda \,x \,(1-\rho)-e\,\rho \,\right]P(\rho,x,t)\right\} &
\\
 -  \, \partial_{x}\left \{ \left[\,c \,h(f) \,\rho - \lambda\, x \,\right] P(\rho,x,t)\right\}
 &\\
 + \frac{1}{2M}\, \partial_{\rho}^2 \left \{ \left[\,e\,\rho+\lambda \,x\, (1-\rho) \,\right]P(\rho,x,t)\right\} 
 &\\
 + \frac{1}{2M}\, \partial_{x}^2 \left \{ \left[\,c \,h(f) \,\rho+ (\lambda + 2 \,\mathcal{D}) \,x \,\right]P(\rho,x,t)\right\} 
 &\\
 - \frac{1}{M}\, \partial_{x}  \partial_{\rho} \left[\,\lambda\, x \,(1-\rho)\,P(\rho,x,t)  \right]
\end{split}
\end{equation}
as shown in Appendix \prettyref{app:mf}.
\prettyref{eq:MFFP} is a two-dimensional Fokker-Planck equation in the representative variables $\rho$ and $x$.
Hence, the corresponding Langevin equation can be straightforwardly derived, adopting the It\^o prescription:
\begin{equation}\label{eq:langevinMF}
    \hspace{-0.35cm} 
    \begin{dcases*}
       \dot{x} =
       c \,h(f) \,\rho - \lambda \, x 
       +\mathbb{B}^{x x}\,\eta_{x} 
       \\
        \dot{\rho} = 
         \lambda \,x \,(1-\rho)- e\,\rho
       + \mathbb{B}^{\rho\rho}\,\eta_{\rho} +
       \mathbb{B}^{\rho x}\,\eta_{x}
   \end{dcases*}
\end{equation}
where $\vec{\eta}=( \eta_\rho,\eta_x)$ are uncorrelated white noises and the two-dimensional matrix $\mathbb{B}$ is related to the Fokker-Planck diffusion matrix $\mathbb{\hat{D}}$ of \prettyref{eq:MFFP} through $\mathbb{\hat{D}} =  \mathbb{B}\mathbb{B}^T$:
\begin{align}
   \mathbb{\hat{D}}
    = \frac{1}{M}
    \begin{pmatrix}
        c \,h(f) \,\rho+(\lambda+2\,\mathcal{D}) x 
        & - \lambda x  (1-\rho) \\
        - \lambda x  (1-\rho) 
        & e\rho+\lambda x (1-\rho) 
    \end{pmatrix} \,.
\end{align}
Since we are primarily interested in the persistence of settled populations, we further assume a separation of timescales in the dynamics of the explorers and the settled population. This results in a one-variable effective Langevin equation for the settled population (\ref{eq:QSSdet}), mirroring the deterministic generalization of the Hanski and Ovaskainen metapopulation model derived in \cite{metapop}. The elimination of the variable $x$ is, however, more involved than in the deterministic case, due to the presence of noise. We report the detailed calculations in \mbox{Appendix \prettyref{app:mf}}, resulting in the following one-dimensional It\^o SDE:
\begin{align}\label{eq:MFeq}
     \dot{\rho} &=  \left[c \,h(f)\, (1-\rho) -e\right] \rho \,+
     \\ &
    + \sqrt{ 
        \frac{\rho}{M}
    \left\{
        e + \frac{c h(f)}{2(1+\mathcal{D}/\lambda)}
        \left[
            1-\rho^2  + 2 \frac{\mathcal{D}}{\lambda} (1-\rho)
        \right]
    \right\}
        } \, \sigma(t)
    \nonumber
\end{align}
with $\langle \sigma (t) \rangle = 0,\, \langle \sigma (t)\sigma (s) \rangle = \delta(t-s)$. 
To verify our derivation and assumptions, we numerically compare the trajectories resulting from the stochastic QSS equation \prettyref{eq:MFeq} with those obtained from a Gillespie simulation of the complete Master equation, which samples the exact microscopic dynamics of the model described in \prettyref{eq:reactions}. As shown in \mbox{\prettyref{fig:gillespie}}, we achieve a very good matching between the different trajectories in the survival regime. In the extinction regime, the mean densities are similar. However, the Gillespie trajectory reaches extinction more slowly due to the absorbing boundary, which causes the explorer densities to always deviate from the quasi-stationary value.
This suggests that the simplification methods considered up to this point are still able to reproduce the average behavior of the stochastic model under different parameters.

We can now exploit \prettyref{eq:MFeq} to analyze the stochastic dynamics of the model. In particular, the mean-field settled population density $\rho$ has domain $(0,1]$, with an absorbing boundary at $\rho=0$, that corresponds to population extinction. $\rho=1$ is instead a reflecting boundary, corresponding to the maximum number $M$  of individuals that each patch can sustain. Deterministically, \prettyref{eq:MFeq} yields an effective metapopulation capacity $\lambda_M = h(f)$,
which still allows us to distinguish two regimes. When $e/c>\lambda_M$, the deterministic stationary solution is $\rho_\mathrm{st} = 0$, corresponding to widespread extinction. If $e/c<\lambda_M$ the stationary solution is instead given by $\rho_\mathrm{st} = \frac{1}{\lambda_M}(\lambda_M - \frac{e}{c})$, which defines the survival regime in the deterministic mean-field model. However, due to the presence of noise, survival and extinction are no longer deterministic. In particular, if one solves the full stochastic equation in \prettyref{eq:MFeq}, with the specified boundary conditions, the only true stationary solution is $\rho_\mathrm{st}=0$. Indeed, the regime $e/c<\lambda_M$ is now characterized by a long-lasting metastable state in which the population survives (see \prettyref{fig:gillespie}). This reflects the well-known fact that phase transitions appear only in the thermodynamic limit of an infinite system size.
Since the diffusion terms in the Fokker-Planck equations are inversely proportional to the local population size $M$, we anticipate that stochasticity will have a more pronounced impact on systems with a small carrying capacity. Indeed, as the population size decreases, the likelihood of a random fluctuation inducing extinction increases.

To investigate these effects, we need to compute the probabilities of survival and the time taken to reach extinction. 
These quantities provide comprehensive information about the impact of the absorbing boundary and enable us to ascertain whether the regimes defined by the deterministic dynamics remain relevant when stochasticity is considered.


\section{\label{sec:scaling}scaling properties close to the transition}
\noindent From the effective stochastic dynamics given by \prettyref{eq:MFeq}, we can investigate how the system size $M$, which is related to the amount of noise affecting the system, influences the survival probability.

Given an initial condition $\rho$ at time $t=0$, the survival probability is defined as the probability that the variable is still in its domain $(0,1]$ at time $t$ \cite{Gardiner}:
\begin{align}\label{eq:S}
    & S(t| \rho,M) = \int_0^1 \,  p(\rho',t|\rho,0) \,d \rho' \,,
\end{align} 
where $p(\rho',t|\rho,0) $ is the solution of the FP equation related to \prettyref{eq:MFeq}, with the initial condition \mbox{$p(\rho',t=0|\rho,0) = \delta(\rho-\rho')$.}
Then, we introduce the mean exit time and its higher moments,
\begin{align}
  T_n(\rho,M) = \int_0^\infty p(t|\rho) \, t^n\,dt \,,
\end{align}
where $p(t|\rho)$ is the probability distribution of extinction times, given the initial condition $\rho$. The latter probability distribution is connected to the survival probability through $p(t|\rho) = -\partial_t S(t|\rho)$ \cite{Gardiner}, which implies
\begin{align}\label{eq:TnS}
   T_n(\rho,M) =\int_0^\infty \, t^{n-1} S(t|\rho,M)\,dt \,.
\end{align}
To investigate how the survival probability changes with the carrying capacity, we simulate the dynamics of the mean-field QSS stochastic differential equation corresponding to \prettyref{eq:MFeq} for various values of $M$, using the Euler-Maruyama algorithm, with a total of $10^7$ realizations. \prettyref{fig:scaling1}a shows the numerical survival probability curves, with the system set at the boundary between the two regimes identified in the deterministic limit, i.e. $e/c \equiv \lambda_M$.
We observe that the survival probability curves appear to exhibit a power-law behavior.

Hence, we adopt a typical finite-size scaling ansatz for the survival probability:
\begin{align}\label{eq:hyp}
    S(t| \rho,M) \propto \frac{1}{t^{\alpha}}\mathcal{F}_\rho(t \,M^\phi) \; .
\end{align}
This corresponds to assuming that, close to the critical point, the survival probability is a generalized homogeneous function of time $t$ and system size $M$. The scaling function $\mathcal{F}_\rho$ depends on $t$ and $M$ only through their combination $t \,M^\phi$, and the exponents $\phi$ and $\alpha$ are expected to be independent of the details of the system.
Plugging our ansatz (\ref{eq:hyp}) into \prettyref{eq:TnS} yields:
\begin{align}\label{eq:1}
    & T_n(\rho,M)
    = 
    \int_0^\infty \mathcal{F}_\rho(t M^\phi) \,t^{n-\alpha-1} dt 
    = M^{-\phi (n-\alpha)}\, C_{n,\alpha}(\rho) \,.
\end{align}
Notice that, by taking the ratio of two consecutive moments - as detailed in Appendix \prettyref{app:survprob} - we obtain
\begin{align}\label{eq:2}
     \frac{ \langle T^{n+1}\rangle }{  \langle T^n\rangle } =  \frac{ C_{n+1,\alpha}(\rho)}{ C_{n,\alpha}(\rho)}\, M^{-\phi}
\end{align}
where $C_{n,\alpha}(\rho)$ depend only on the initial condition $\rho$. We can exploit the above relations to numerically verify whether the finite-size scaling ansatz is correct and obtain a preliminary numerical estimate of the scaling exponents  $\alpha$ and $\phi$. 
By extracting the time to extinction from each realization of the simulation and fitting the moments' ratios as in \prettyref{eq:2}, we can directly obtain the value of $\phi$ and subsequently estimate $\alpha$ from \prettyref{eq:1}:
\begin{equation}\label{eq:fittedexp}
    \phi =-0.506 \pm 0.002 ,
    \quad 
    \alpha = -0.005 \pm  0.003 \; .
\end{equation}
Further details on the numerical derivation are provided in Appendix \prettyref{app:numexp}.
Ultimately, \prettyref{fig:scaling1}b shows the collapse of the survival probability curves rescaled according to the numerical exponents in \prettyref{eq:fittedexp}, which confirms our scaling hypothesis given in \prettyref{eq:hyp}.

Yet, obtaining the scaling behavior of our system away from criticality presents more challenges: indeed, choosing a suitable scaling ansatz is not straightforward in this case. For this reason, we will adopt a more rigorous approach and analytically derive the scaling form of the moments of extinction time directly from the effective stochastic equation (\ref{eq:1}). This method yields the exact values of the scaling exponents even when the system is not poised at the critical point.

\begin{figure*}[t]
    \centering\includegraphics[width=0.95\linewidth]{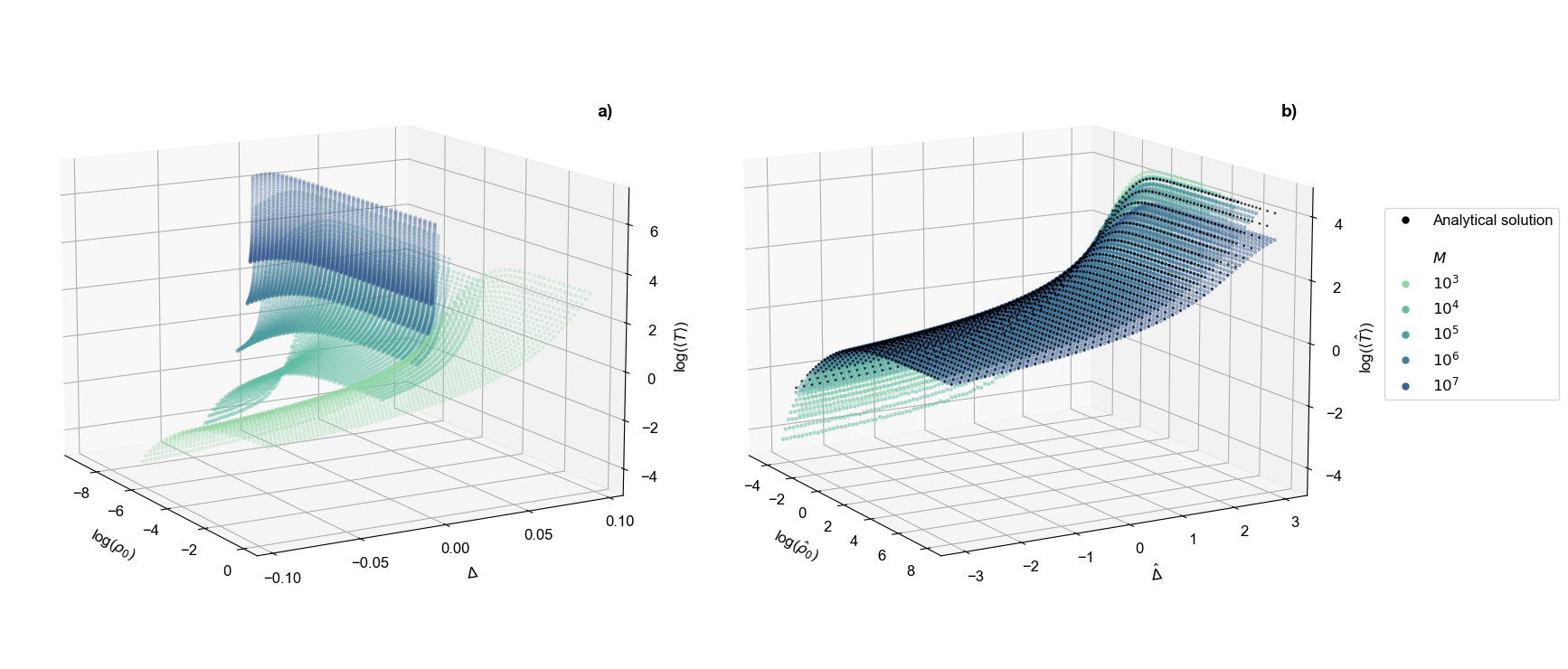} 
    \caption{\raggedright
    First moment of exit time $\langle T \rangle$ versus initial condition $\rho_0$ and deviation from criticality $\Delta$ for different values of the local carrying capacity $M$. \textbf{a)} not rescaled. \textbf{b)} rescaled according to \prettyref{eq:T1}. The black dots indicate the analytical solution \prettyref{eq:asoltocrit}.
    Averages are obtained from $10^5$ numerical realizations of the MF QSS dynamics in \prettyref{eq:MFeq}.}
    \label{fig:scaling}
\end{figure*} 

\subsection{First and second moments of exit time\label{sec:12mom}}
\noindent 
We set $T \equiv T_1$ and analytically investigate the scaling behavior of $T_n$ with the carrying capacity $M$. We rely on the following differential equations, derived from the backward Fokker-Planck equations for the survival probability \cite{Gardiner}:
\begin{align}\label{eq:firstmoment}
    A(\rho)\,\partial_{\rho} T(\rho)+\frac{1}{2}D(\rho)\,\partial_{\rho}^2 T(\rho) = -1
\end{align}
\begin{align}\label{eq:nthmoment}
    &A(\rho)\,\partial_{\rho} T_n(\rho)+\frac{1}{2}D(\rho)\,\partial_{\rho}^2 T_n(\rho) = -n \, T_{n-1}(\rho)
\end{align}
\normalsize
where $A(\rho)$ and $D(\rho)$ are the drift and diffusion coefficients of \prettyref{eq:MFeq}, respectively, and $\rho$ is the initial condition of the settled population density at time $t=0$. \\
We consider a setting in which the system is close to criticality and define the deviation from the critical point ($e/c=\lambda_M $) through
\begin{align}\label{eq:delta}
    \Delta \coloneqq \frac{1}{\lambda_M}\left( \lambda_M - \frac{e}{c}\right) = \frac{\epsilon}{\lambda_M}, \qquad |\epsilon| \ll 1 
\end{align}
where $\lambda_M = h(f)$, as resulting from the deterministic part of the effective quasi-stationary Fokker-Planck \prettyref{eq:MFeq}. 
As shown in Appendix \prettyref{app:1stmom}, by rescaling  $\rho \rightarrow \hat{\rho} = \rho\, \sqrt{M}$, $T \rightarrow \hat{T} = \frac{\lambda_M c}{\sqrt{M}} \,T$ and $\Delta \rightarrow \hat{\Delta} = \Delta \sqrt{M}$, the equation for the first moment of exit time \prettyref{eq:firstmoment} becomes
\begin{align}
\label{eq:closetocrit}
 \left(\hat{\Delta} - \hat{\rho} \right)\hat{\rho}\, \partial_{\hat{\rho}} \hat{T}+\frac{\chi}{4}\hat{\rho} \,\partial_{\hat{\rho}}^2 \hat{T}  = - 1 
\end{align}
in the limit of large system sizes $M$. Here, we have introduced $\chi \coloneqq 3 + \frac{\mathcal{D}}{\mathcal{D}+\lambda}$, which we treat as a fixed constant.
Since \prettyref{eq:closetocrit} is independent of $M$, the rescaled extinction time $\hat{T}$ is expected to be a function of the rescaled initial density $\hat{\rho}$ and the rescaled distance from the critical point $\hat{\Delta}$ alone. This implies that, in the limit of large $M$, $T$ can be written as the scaling form
\begin{align}
\label{eq:T1}
    T(\rho,\Delta,M) \propto \sqrt{M} \, \hat{T} \left( \sqrt{M} \rho, \sqrt{M} \Delta \right)
\end{align}
which implies that it is a homogeneous function of $\rho$ and $\Delta$.
The solution of \prettyref{eq:closetocrit}, with boundary conditions
$\hat{T}(0) = 0 $,
$\partial_{\hat{\rho}}\hat{T}(\hat{\rho}) \rvert_{\hat{\rho} = \sqrt{M}} =  0 $,
is given by the integral \prettyref{eq:asoltocrit}, which can be evaluated analytically at criticality ($\hat{\Delta}$=0) and numerically elsewhere. 
To test our analytical predictions, we simulate the mean-field QSS SDE over $10^5$ realizations, varying the distance from the critical point $\Delta$, the initial condition in the settled population density $\rho$, and the system size $M$.
As shown in \prettyref{fig:scaling}a, the simulated points distribute within a volume in the three-dimensional space defined by $\Delta$, $\rho$, and $T$. Upon rescaling the axes with $M$ according to \prettyref{eq:T1}, the points collapse onto a single well-defined surface that encompasses different initial conditions and distances from criticality, consistently with the numerical evaluation of the analytically derived average extinction time (\prettyref{eq:asoltocrit}).
The effectiveness of the collapse can be more clearly visualized through the two-dimensional sections plotted in the left panel of \prettyref{fig:slices}.
We can now apply the same procedure to the equation for the second moment of the exit time, obtained by substituting $n=2$ into \prettyref{eq:nthmoment}. The same analysis yields the scaling expression
\begin{align}\label{eq:T2} 
   T_2(\rho,\Delta,M) \propto   M \, \hat{T}_2\left( \sqrt{M} \rho, \sqrt{M} \Delta \right) \,.
\end{align}
\prettyref{fig:scalingT2} shows that rescaling the surfaces obtained from simulation data according to \prettyref{eq:T2} leads to the collapse of these surfaces. We present further details of the derivation, along with an explicit analytical solution of the differential equation for  $T_2$, in Appendix \prettyref{app:1stmom}.

\subsection{Survival probability}
\noindent Having found $\langle T \rangle \propto \sqrt{M}$ (\prettyref{eq:T1}) and $\langle T^2 \rangle  \propto M$ (\prettyref{eq:T2}), we can determine the scaling of the survival probability at criticality ($\Delta$ = 0). The $\phi$ exponent is obtained from \prettyref{eq:2} as
\begin{align}
    \frac{\langle T^2 \rangle}{ \langle T \rangle} \propto \sqrt{M} 
    \quad \Rightarrow \quad  \phi \equiv -\frac{1}{2}, 
\end{align}
and, similarly, we can calculate the $\alpha$ exponent (\ref{eq:1}) as
\begin{align}
     \langle T \rangle \propto  M^{-\phi (1-\alpha)}
     \quad \Rightarrow \quad \alpha=0
\end{align}
which yields the scaling form
\begin{align}\label{eq:scalingS}
    S(t|M) \propto f(t \,M^{-\frac{1}{2}}) \; .
\end{align}
It is worth noting that the scaling exponents derived analytically and the ones obtained numerically reported in \prettyref{eq:fittedexp} agree remarkably well.

The technique we employed to analyze moments of extinction time can also be replicated for the survival probability. The starting point is once more the backward Fokker-Planck equation for the survival probability \cite{Gardiner},
\begin{align}\label{eq:survprob}
  \partial_t S &= A(\rho)\,\partial_{\rho} S+\frac{1}{2}D(\rho)\,\partial_{\rho}^2 S \; .
\end{align}
Since \prettyref{eq:survprob} involves a time derivative, it is necessary to rescale both the initial condition and the time as
\begin{align}
    & \rho \rightarrow \hat{\rho}= \rho\, M^{\gamma}
    \qquad t \rightarrow \hat{t}=t \,\lambda_M\, c \, M^\phi \; .
\end{align}
After setting $\phi=-\frac{1}{2}$ and $\gamma=\frac{1}{2}$, we obtain the size-independent equation
\begin{align}
    \partial_{\hat{t}} S = 
    (\hat{\Delta} - \hat{\rho})  \hat{\rho} \,\partial_{\hat{\rho}} S
    +\frac{\chi}{4} \hat{\rho} \,\partial{\hat{\rho}}^2 S
\end{align}
from which we derive
\begin{align}
   S = S\left( \frac{t}{\sqrt{M}}, \sqrt{M}\rho, \sqrt{M}\Delta \right)
\end{align}
as detailed in Appendix \prettyref{app:survprob}. This scaling is consistent with our previous findings regarding the first and second moments of extinction time, but it now offers a more general form that relates to the entire distribution of extinction times rather than its specific moments.
By fixing the distance from the critical point $\Delta=0$ and the initial condition $\rho$, we recover our result in \prettyref{eq:scalingS}.

Overall, the scaling relations of survival probability at criticality and moments of extinction time away from criticality can be summarized as follows:
\begin{align}
    \begin{split}
        &  S(t|M) = t^{-\alpha} \hat{S}( t M^{\phi})
        \\
        & T_1(\rho,\Delta, M)= M^{\beta} \hat{T}_1 (\rho M^{\gamma}, \Delta M^{\eta})
        \\
        & T_2(\rho,\Delta, M)= M^{\delta} \hat{T}_2 (\rho M^{\gamma}, \Delta M^{\eta})
    \end{split}
\end{align}
with the exponents connected through the scaling relations
\begin{align}
    \beta=-\phi(1-\alpha),  \quad  \delta=-\phi(2-\alpha) \; .
\end{align} 
Their numerical values are reported in \prettyref{tab:scaling}.

\begin{table}[h]
\begin{ruledtabular}
\begin{tabular}{cccccc}
$\alpha$ 
&$\phi$ 
&$\gamma$ 
&$\eta$ 
&$\beta$\normalsize 
&$\delta$\normalsize
\rule[-2ex]{0pt}{5ex} \\
\hline 
0 & $-\frac{1}{2}$ &$\frac{1}{2}$ &$\frac{1}{2}$
&$\frac{1}{2}$ &1
\rule[-2ex]{0pt}{5ex}
\rule[-2ex]{0pt}{5ex} 
\end{tabular}
\end{ruledtabular}
\caption{
    \raggedright
    Scaling exponents characterizing the survival probability $S$ and the first and second moments of the exit time, $T_1$ and $T_2$ in the vicinity of the critical point.
    }
\label{tab:scaling}
\end{table}

\section{Coarse-grained simplifications for ecological directions}

\begin{figure*}[ht!]
    \centering
    \includegraphics[width=0.8\textwidth]{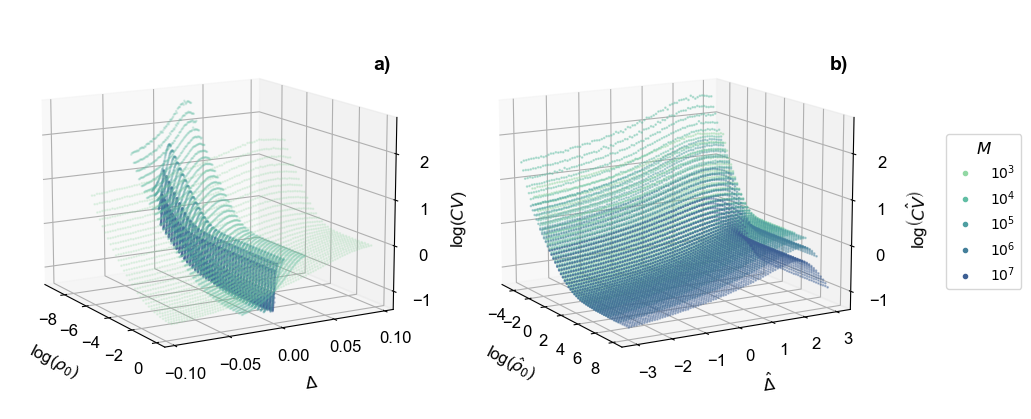}
    \vspace{1em}
    \hspace{-3em}
        \centering
        \includegraphics[width=0.68\textwidth]{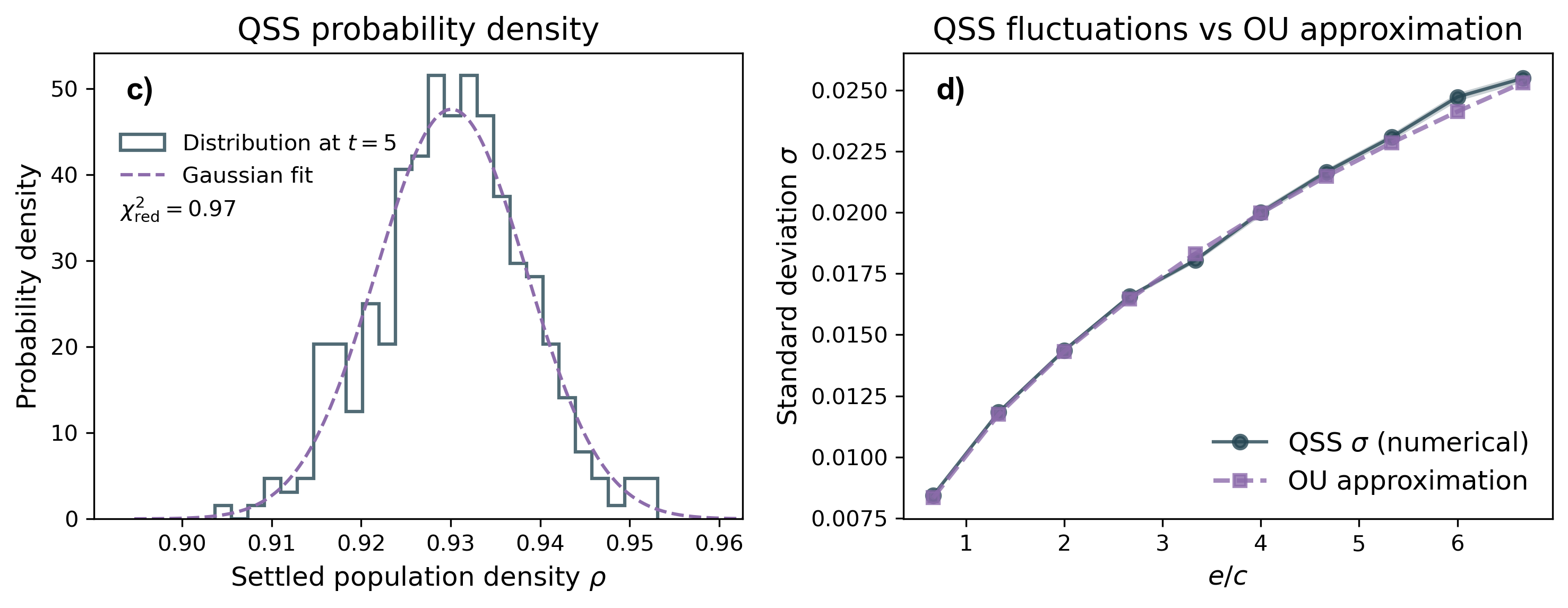}
    
    \caption{
    \raggedright
    \textbf{Top:} Coefficient of variation (CV) versus initial condition $\rho_0$ and deviation from criticality $\Delta$ for different values of the local carrying capacity $M$. \textbf{a)} not rescaled. \textbf{b)} rescaled according to \prettyref{eq:T1} and \prettyref{eq:T2}. 
    Averages are obtained from $10^5$ numerical realizations of the QSS dynamics in \prettyref{eq:MFeq}.
    \textbf{Bottom:}  Fluctuations in the survival regime of the QSS dynamics given by \prettyref{eq:MFeq}.
    \textbf{c)}  Probability distribution of the settled population density at \(t \approx 5\), along with a Gaussian fit. The same parameters are used as in the survival trajectories shown in \prettyref{fig:gillespie}, and the distribution is obtained from 350 independent realizations. The reduced chi-squared of the fit is \(\chi^2_{\mathrm{red}} = 0.97\).
    \textbf{d)} Standard deviation \(\sigma\) as a function of \(e/c\), comparing results from direct numerical integration of the QSS dynamics with the linearized analytical estimate based on the Ornstein--Uhlenbeck (OU) approximation. The numerical values of \(\sigma\) are obtained from Gaussian fits to distributions built from 250 independent realizations.
    }
\label{fig:eco}
\end{figure*}
\noindent 
Thus far, our analysis has considered a mean-field setting with a fully connected dispersal network.
While analytically tractable, these assumptions are typically unrealistic in natural ecosystems. 
In this section, we develop tools based on mean-field framework that can 
nonetheless yield ecologically meaningful insights.

We first consider the coefficient of variation (CV) of the time to extinction, defined as the ratio of the standard deviation to the mean i.e., $CV(T) =\sqrt{\langle T^2 \rangle - \langle T \rangle^2}/\langle T \rangle$.  This estimate, which captures relative fluctuations, provides a way to quantify the uncertainty and sensitivity of ecological time series. 
Figures \ref{fig:eco}a and b show that the CV exhibits a collapse behavior similar to that of the moments, as expected. Interestingly, we also observe the emergence of an optimal value: increasing the rescaled distance from criticality, $\hat \Delta$, causes the CV to rise, reaching a maximum at $\hat \Delta > 0$. We recall that $\hat \Delta = 0$ corresponds to the critical point separating extinction and survival regimes, which are defined for the deterministic dynamics obtained for $M\to \infty$. Here, we see that a finite $M$ may lead to an additional transition point, associated with the variability in the time to extinction, namely the point at which the CV attains its maximum.
This behavior is further highlighted in the two-dimensional slices of the CV collapse in \mbox{\prettyref{fig:CVslices}.} This transition point warrants further investigation, especially in more realistic networks for which the coefficient of variation is obtainable from time-series data. 

As observed in \prettyref{fig:gillespie}, once the system transitions into the survival regime, it enters a metastable state in which the settled population fluctuates around a well-defined mean. The corresponding quasi-stationary dynamics, described by the stochastic differential equation in \prettyref{eq:MFeq}, involves nonlinear drift and a diffusion term with higher-order polynomial dependence.
To simplify this description, we ask whether the fluctuations near the deterministic metastable state can be approximated by a linearized, effective process. This would enable analytical predictions about variability and responses to environmental perturbations—potentially applicable beyond the fully connected mean-field case \cite{PhysRevE.101.062132}.
To this end, we consider an Ornstein–Uhlenbeck (OU) approximation to the stochastic dynamics. Neglecting the absorbing boundary, this leads to a Gaussian approximation of the stationary distribution. For the deterministic dynamics, \prettyref{eq:MFeq}
admits a stable survival solution at $\rho_* = 1-\frac{e}{c\, h(f)}$. Thus, linearizing the drift term and evaluating the noise term at $\rho_*$ yields the effective drift rate $\kappa$ and diffusion coefficient $D_{\rm eff}$:
\begin{align}
    &\kappa = \bigg\rvert \frac{\partial}{\partial \rho} A(\rho)\big\rvert_{\rho_*}\bigg\rvert = |e - c\, h(f)| 
    \nonumber \\
     &D_{\rm eff} = D(\rho_*) = 
     \nonumber\\
     & \qquad =\frac{\rho_*}{M}\left\{e+\frac{c \,h(f)}{2(1+\mathcal{D}/\lambda)}\left[1-\rho^2_* + 2 \frac{\mathcal{D} }{\lambda} (1-\rho_*) \right]\right\} 
     \nonumber\\
     & \qquad =\frac{1}{2M} \frac{e (e-c h(f)) (e \lambda -4 c \,h(f) (\mathcal{D}+\lambda ))}{(c\, h(f))^2  (\mathcal{D}+\lambda )}
    \nonumber \\
  & \sigma_{\rm eff} = \sqrt{\frac{D_{eff}}{2\kappa}}
\end{align}
and the resulting linearized dynamics is governed by the effective OU equation:
\begin{align}
    \dot \rho_{\rm OU} = - \kappa \,\rho_{\rm OU} + \sqrt{2 \sigma_{\rm eff}} \,\eta(t)
\end{align}
where $\eta(t)$ is gaussian white noise. To evaluate the validity of this approximation, we fit the stationary settled densities to a Gaussian distribution and compare the resulting standard deviation with that predicted by the OU process with the effective drift and diffusion coefficients.
As shown in \prettyref{fig:eco}c,  the empirical distribution in the metastable state is well approximated by a Gaussian.
In \prettyref{fig:eco}d, we compare the standard deviation predicted by the OU approximation to that obtained from Gaussian fits across a range of parameter ratios $e/c$, finding excellent agreement. This indicates a potential simplification of the metastable dynamics, which may hold beyond our simplistic model assumptions. 

\vspace{2em}
\section{Conclusions}

\noindent In this work, we extended the traditional metapopulation framework to incorporate stochasticity, with the aim of investigating its impact on metapopulation persistence. 
We focused on understanding the role of finite carrying capacity in the subpopulation size, which is directly associated with the strength of demographic fluctuations. In doing so, we were able to quantitatively address the dynamics of finite metapopulations and study their stochastic evolution. 

Our results suggest that, for a simple colonization-death dynamics in homogeneous landscapes governed by mean-field dispersal networks, the introduction of stochasticity does not lead to strong deviations from the deterministic population dynamics. Yet, we uncovered the unexpected appearance of an absorbing boundary even in the regime in which the population would deterministically survive. Hence, the stochastic metapopulation always faces eventual extinction due to the finite carrying capacity. Additionally, we have found strong consistency with finite-size scaling predictions, as evidenced by the collapse of the survival probability and extinction time moments curves. The scaling exponents governing this collapse have been analytically derived from the backward Fokker-Planck equation of the model (\prettyref{tab:scaling}) as well as verified through numerical simulations. These results provide a baseline expectation for further studies into stochastic metapopulation models. The bottom-up derivation of the Fokker-Planck equation and the effective Langevin equation present immediate numerical applications beyond mean-field networks to incorporate interesting topological factors which we omitted for the sake of analytical simplicity. Changes in the scaling exponents will not only provide possible connections to universality classes but also connect network topology and average time to extinction induced by demographic stochasticity.

Despite the fact that natural ecosystems are far more complex than the simplified scenario considered here, our results provide a quantitative link between a system's proximity to the extinction threshold, as well as both the expected time to extinction and the variability around that time. In particular, we have shown that survival time increases sharply as the system moves away from the critical point, while relative fluctuations (as measured by the coefficient of variation) become smaller only after a transition point that is higher than the metapopulation capacity. This relationship is especially useful when independent estimates of the deterministic metapopulation capacity are available (e.g., from connectivity measures or colonization-extinction data) alongside the local carrying capacity. Our results demonstrate how these two parameters alone can offer meaningful predictions about extinction time statistics in finite systems both above and below the critical value. This aspect is particularly significant because the classic metapopulation capacity does not carry any information at all regarding timescales. These timescales can be a proxy for the robustness to fluctuation-induced extinction events. In this respect, we emphasize that the noise term we derived, which can cause extinctions also in the deterministic persistence scenario, goes beyond the usual demographic noise from pure birth-and-death processes, incorporating the effective contribution of colonization dynamics. Non-trivial network topologies are also expected to play an important role in shaping the fluctuations of the metapopulation density. For example, dendritic river networks affect ecological dynamics \cite{carrara2012dendritic,nauta}, and the spatial structure of the diffusion network can influence the probability of fixation of a mutation in stochastic metapopulation models of microbial communities \cite{MarrecBitbol2021,AbbaraBitbol2024}. Similarly, environmental cofactors, such as wildfires that influence forest dynamics \cite{nicoletti2023emergence}, and stochastic or time-varying dispersal \cite{williams2013stochastic, snyder2014much, martin2020intermittent}, can significantly alter the impact of demographic stochasticity. Our effective SDE provides a framework to capture these effects by appropriately modeling boundary conditions \cite{BerLin2022,Su_2023,BookChap2024}. Given a simplified dynamics of the metastable state, we can also expect to understand the effect of ecologically relevant networks through their distance from the effective drift and diffusion parameters. In time series data, where local carrying capacity information is also available, more involved methods can be used to infer the drift and diffusion terms \cite{nabeel2025discovering} which may enable the quantification of the distance from mean-field connectivity. Such methods could be used as indirect measures of the connectivity of a landscape. The existence of long metastable states also facilitates the study of the response to environmental perturbations on both dynamics of population and its persistence \cite{padmanabha2023generalisation}. In this sense, demographic noise is not merely a complicating factor — it carries early-warning information about system stability. 

Finally, a key insight from our analysis is the universality of the scaling behavior. The mathematical structure governing the relationship between capacity, fluctuations, and extinction time does not depend on the fine details of the system. As a result, data collected in one habitat could be used to predict survival time statistics in another, provided the key parameters—capacity and carrying size—are known. This universality implies practical value for the theory, enabling better generalization across species or landscapes without requiring full system-specific modeling each time. An exciting direction for future work is to investigate whether this universality holds beyond the mean-field approach pursued here.

Overall, our results provide a foundation for integrating more complex and realistic effects into stochastic metapopulation models. By refining microscopic processes and dispersal network topology, these extensions will further elucidate the critical role of stochasticity in the survival of spatially structured populations.

\vspace{2em}

\noindent \textbf{Code availability} \\
All the codes and simulation data used to generate the figures are available at \cite{code_availability}


\begin{acknowledgments}
\noindent
A.D. and D.B. were supported by the Italian Ministry of University and Research (project funded under the National Recovery and Resilience Plan (NRRP), Mission 4, Component 2 Investment 1.4 - Call for tender No. 3138 of 16 December 2021, rectified by Decree n.3175 of 18 December 2021 of Italian Ministry of University and Research funded by the European Union – NextGenerationEU; Award Number: Project code CN\_00000033, Concession Decree No. 1034 of 17 June 2022 adopted by the Italian Ministry of University and Research, CUP C93C22002810006, Project title \quotes{National Biodiversity Future Center - NBFC}). G.N. acknowledges funding provided by the Swiss National Science Foundation through its Grant CRSII5\_186422. P.P. was supported by NCCR Microbiomes Grant 51NF40\_180575 from the Swiss National Science Foundation. The authors thank Amos Maritan and Sandro Azaele for very helpful discussions and important comments on the paper.
\end{acknowledgments}

\appendix

\section{Step-operator form of the master equation}\label{app:step_operators}
\noindent
To facilitate the system size expansion, we reformulate the master equation \prettyref{eq:ME} using step operators, which shift occupation numbers in the state vector \( \vec{n} \).
The step operator \( \xi_{A_i}^{\pm1} \) acts on a generic function \( f(\vec{n}) \) as follows:
\begin{equation}
\xi^{\pm1}_{A_i} 
f(n_{A_1},... n_{A_i},... n_{A_N}) = 
f(n_{A_1},... n_{A_i}\pm1,... n_{A_N}) 
\end{equation} 
Using this notation, the master equation can be rewritten in a more compact form \prettyref{eq:master}. 
\\
For example, consider the gain term associated with a process in which a settled individual at patch \(i\) dies:
\begin{equation}
W(n_{S_i}; n_{X_i} | n_{S_i}+1; n_{X_i}) \, \mathcal{P}(n_{S_i}+1; n_{X_i}) \;.
\end{equation} 
This term represents the contribution to the probability \(\mathcal{P}(n_{S_i}, n_{X_i})\) due to transitions from state \((n_{S_i} + 1, n_{X_i})\) to the state \((n_{S_i}, n_{X_i})\). 
Applying the step operator \(\xi^{+1}_{S_i}\), it can be rewritten as:
\begin{equation}
\begin{aligned}
&
W(n_{S_i}; n_{X_i} | n_{S_i}+1; n_{X_i}) \, \mathcal{P}(n_{S_i}+1; n_{X_i})  = \\
&  = \xi_{S_i}^{+1}
 W(n_{S_i} - 1; n_{X_i} | n_{S_i}; n_{X_i})  \, \mathcal{P}(n_{S_i}; n_{X_i}) \,.
\end{aligned}
\end{equation}
At the same time, this process under consideration also yields a loss contribution to the probability $\mathcal{P}(n_{S_i}; n_{X_i}) $:
\[
W(n_{S_i}-1; n_{X_i} | n_{S_i}; n_{X_i}) \, \mathcal{P}(n_{S_i}; n_{X_i}) \;.
\]
As a result, the net effect of the settled individual's death process can be compactly expressed as:
\begin{equation}
\begin{aligned}
\left( \xi_{S_i}^{+1} -1 \right)
 W(n_{S_i} - 1; n_{X_i} | n_{S_i}; n_{X_i})  \, \mathcal{P}(n_{S_i}; n_{X_i}) \,.
\end{aligned}
\end{equation}
Similarly, when multiple variables are involved, a compound operator like \( \xi_{S_i}^{-1} \xi_{X_i}^{+1} \) acts by shifting both indices:
\small
\begin{equation}
\begin{aligned}
&\left( \xi_{S_i}^{-1} \xi_{X_i}^{+1} - 1 \right) W(n_{S_i} + 1; n_{X_i} - 1 | n_{S_i}; n_{X_i}) \, \mathcal{P}(n_{S_i}; n_{X_i}) = \\
& \quad = W(n_{S_i}; n_{X_i} | n_{S_i} - 1; n_{X_i} + 1) \, \mathcal{P}(n_{S_i} - 1; n_{X_i} + 1)  \, + \\
& \qquad 
- W(n_{S_i} + 1; n_{X_i} - 1 | n_{S_i}; n_{X_i}) \, \mathcal{P}(n_{S_i}; n_{X_i}) \;.
\end{aligned}
\end{equation}
\normalsize
Applying this notation to all the processes with their respective transition rates (\prettyref{eq:Wrates}), the master equation takes the form given in \prettyref{eq:master} of the main text. 
\\
To maintain a compact and readable form, we have grouped the probability distribution \( \mathcal{P} \) at the end of each term. 
However, it is important to note that the step operators act on all functions appearing to their right - that is, the full product \( W \mathcal{P} \), with the arguments of both the transition rates and the probability being shifted accordingly.



\section{Mean-field model\label{app:mf}}
\noindent
Let us consider a complete network (whose adjacency matrix is $\mathcal{A}_{ij} = \mathcal{A}\, (1-\delta_{ij})$ with  $\mathcal{A}=1/N$) and assume patch homogeneity: $c_i = c \,\, \forall \, i, \quad e_i = e \,\, \forall \, i $. 
Under these hypotheses, the drift \prettyref{eq:drift} and diffusion \prettyref{eq:diffcomponents} coefficients of the Fokker-Planck equation become:
\small
\begin{align}\label{eq:fullyconn}
    & \hspace{-0.25cm} 
    A^{\rho}_i =   \lambda \,x_i (1-\rho_i) -  e\, \rho_i  \nonumber \\ & \hspace{-0.25cm}
    A^{x}_i =  
     \frac{1}{N}\sum_j \Big( \mathcal{D} x_j
    +c h(f) \rho_j \Big) 
    - \frac{1}{N}c h(f) \rho_i   - (\mathcal{D}+\lambda) x_i
     \nonumber \\ & \hspace{-0.25cm}
    \mathbb{D}^{\rho\rho}_{ij} = \frac{1}{M}\left[ e \rho_i + \lambda  x_i (1 - \rho_i) \right]\delta _{ij}
    \nonumber \\ & \hspace{-0.25cm}
    \mathbb{D}^{\rho x}_{ij} =   \mathbb{D}^{x \rho}_{ij} = -\frac{1}{M} \lambda x_i (1-\rho _i) \, \delta _{ij}  
    \\ & \hspace{-0.25cm}
   \mathbb{D}^{xx}_{ij}  = 
    \frac{1}{M}\Big[ \frac{1}{N}\sum_k  \Big(\mathcal{D}x_k+  c h(f) \rho_k \Big) \,+
     \nonumber \\ & \hspace{-0.55cm} \qquad \qquad
    + (\mathcal{D}+\lambda) x_i  
    - \frac{1}{N} c h(f) \rho_i  \Big]\,\delta_{ij}
    - \frac{2}{MN} \mathcal{D} x_i (1-\delta_{ij})
    \nonumber
\end{align}
\normalsize
After applying the marginalization procedure described in \prettyref{eq:margP} and (\ref{eq:margappr}), we obtain a two-dimensional Fokker-Planck equation  in the representative variables $(x,\rho)$, as reported in \prettyref{eq:MFFP}. The drift vector and diffusion matrix are given by:
\begin{align}
& \vec{\hat{A}}=
\begin{pmatrix}
    \lambda \,x \,(1-\rho)- e\,\rho\\
    c \,h(f) \,\rho -\lambda\, x 
\end{pmatrix}
\nonumber \\
& \mathbb{\hat{D}} =
 \frac{1}{M} 
\begin{pmatrix}
     c \,h(f) \rho +(\lambda+2\mathcal{D}) x 
    & - \lambda x (1-\rho) 
    \\
    - \lambda x (1-\rho) 
   &
   e\rho +\lambda x (1-\rho) 
\end{pmatrix}
\end{align}
The corresponding Langevin equation in \prettyref{eq:langevinMF} can be derived by imposing $\mathbb{B}\mathbb{B}^T = \mathbb{\hat{D}}$ for $\mathbb{B}$ in the form:
\begin{align}
& \mathbb{B} =
\begin{pmatrix}
    \mathbb{B}^{xx}
    & 0
    \\
    \mathbb{B}^{\rho x}
   & \mathbb{B}^{\rho \rho}
\end{pmatrix}
\end{align}
which yields:
\small
\begin{align}
    \mathbb{B}^{xx} & = \frac{1}{\sqrt{M}} \sqrt{ c \,h(f)\rho +(\lambda+2\mathcal{D})x }
    \nonumber \\
    \mathbb{B}^{\rho x} & = - \frac{1}{\sqrt{M}} \frac{ \lambda x (1-\rho) }{ \sqrt{ c \,h(f) \rho +(\lambda+2 \mathcal{D}) x }}
    \nonumber \\
    \mathbb{B}^{\rho \rho} & = \frac{1}{\sqrt{M}} \sqrt{ e\rho +\lambda x (1-\rho)  - \frac{\lambda^2 x^2 (1-\rho)^2}{c \,h(f) \rho +(\lambda+2\mathcal{D})x } }
\end{align}
\normalsize
\noindent
We can now derive the effective equation in the settled population density, by assuming a separation of timescales in the settled and explorer population's dynamics.
Consider  \prettyref{eq:langevinMF} and let $\bar{x}$ be the stationary value of $x$ such that $\dot{x} = 0$:
\begin{align}
    \bar{x} = \frac{c\,h(f)\,\rho}{\lambda}, \quad \eta_x \equiv 0
\end{align}
If we condition $\rho$ on $x$ being fixed at $\bar{x}$, we obtain:
\begin{align}
    &\dot{\rho} = c \, h(f)\, \rho (1-\rho) - e \, \rho + B^{\rho \rho}\big\rvert_{x=\bar{x}} \eta_\rho
\end{align}
which yields the quasi-stationary stochastic \prettyref{eq:MFeq}.


\section{Numerical scaling exponents for the survival probability}\label{app:numexp}
\noindent
Given the finite size scaling assumption \prettyref{eq:hyp} for the survival probability's time dependence, we can express the $n$-th moment of exit time as:
\small
\begin{align}\label{eq:TnS2}
     \langle T^n (\rho,M)\rangle & = - \int_0^\infty  \partial_t S(t|\rho,M)\,t^{n} \,dt 
     \nonumber\\ &
    =\int_0^\infty S(t|\rho,M)\, t^{n-1} \,dt
\end{align}
\normalsize
by substituting the scaling hypothesis \prettyref{eq:hyp} we obtain:
\begin{align}\label{eq:alphafit}
    \langle T^n (\rho,M) \rangle
    & \propto \int_0^\infty \mathcal{F}_\rho(t M^\phi) t^{n-\alpha-1} dt 
    \nonumber \\ & 
    = M^{-\phi (n-\alpha)} \int_0^\infty \mathcal{F}_\rho(z) z^{n-\alpha-1} dz 
    \nonumber \\ & 
    = M^{-\phi (n-\alpha)}\, C_{n,\alpha} (\rho)
\end{align}
and 
\begin{align}\label{eq:phifit}
    \frac{ \langle T^{n+1}\rangle}{ \langle T^n\rangle} =  \frac{ C_{n+1,\alpha}(\rho)}{ C_{n,\alpha}(\rho)} M^{-\phi}
\end{align}
where the explicit dependence on the initial condition $\rho$ is included in the $C_{n,\alpha}$ coefficients.
The above \prettyref{eq:phifit} allows to numerically estimate the $\phi$ exponent. To achieve this, we simulate $10^7$ realizations of the MF QSS dynamics given by \prettyref{eq:MFeq} and we compute the moments of extinction time. Specifically, we select the moments of order $n$ ranging from 1 to 8, and consider 20 values for $M$, spanning from $10^3$ to $10^7$. For any two consecutive moments, we fit the logarithm of $\frac{ \langle T^{n+1} \rangle}{ \langle T^n \rangle}$  as a function of $\log(M)$. In principle, this yields 7 parallel lines. The average of the obtained slopes provides therefore our estimate for $\phi$. The error is taken to be the standard deviation of the slopes.
\\\indent
Once $\phi$ is determined, we can exploit \prettyref{eq:alphafit} to estimate $\alpha$. In particular, we now fit the logarithm of $\langle T^n \rangle$ versus  $\log(M)$, and $\alpha$ is obtained from the slope $m$ of the fitted lines, namely: $m=-\phi (n-\alpha)$. We consider the standard deviation of  $\alpha$  as its error.
\\\indent
However, the estimates of $\phi$ and $\alpha$ are highly influenced from the values of $M$ considered, the number of realizations of the dynamics and the order of the moments used for the fitting. In particular, higher values of $M$ and a larger number of realizations provide more rigorous estimates, as confirmed by deviations from the linearly expected behavior becoming less pronounced. The errors computed as standard deviations are therefore not fully representative of the true uncertainty in the estimates, as they do not account for these factors, which introduce additional variability.


\section{Scaling analysis}
\noindent
Let us express the mean-field QSS stochastic equation
\prettyref{eq:MFeq} in the more convenient Fokker-Planck formalism:
\begin{align}
    &\partial_t P(\rho,t) = -\partial_\rho [A(\rho) P(\rho,t)] + \frac{1}{2}\partial_\rho^2 [D(\rho) P(\rho,t)]
\end{align}
This is a one-dimensional time-homogeneous FP equation. 
The domain of $\rho$ is given by the interval $(0,1]$, with $\rho=0$ absorbing boundary and $\rho=1$ reflecting boundary, as seen in \mbox{\prettyref{sec:QSS}.}
\noindent
Therefore, the definition of survival probability given in \prettyref{eq:S} implies that it obeys the boundary conditions 
$
S(0,t) =0 , \, \partial_\rho S(\rho,t) \rvert_{\rho=1} =0
$. 
Through \prettyref{eq:TnS} we similarly find
$  T_n(0,t) =0 , \, \partial_\rho T_n(\rho,t) \rvert_{\rho=1} =0 
$ for the $n-th$ moment of extinction time.
\\ \indent
For the upcoming calculations, we rewrite the drift and diffusion coefficients of the QSS mean-field Fokker-Planck equation in terms of the parameter $\Delta$  \prettyref{eq:delta}, which quantifies the deviation from criticality:
\begin{align}
    & A(\rho, M, \Delta) = \lambda_M \,c \left( \Delta - \rho \right) \rho \\
    & D(\rho, M, \Delta) = 
     \frac{\lambda_M \, c}{2M}
    \, \rho \,
    \big[ 
    3 + \frac{\mathcal{D}}{\mathcal{D}+\lambda}
    - 2 \Delta 
    \,+ \nonumber\\ &\qquad \qquad \qquad \qquad \quad \quad
    - \frac{2\mathcal{D}}{\mathcal{D}+\lambda}  \,\rho
    - \frac{\lambda}{\mathcal{D}+\lambda}  \rho^2 
    \big] \nonumber
\end{align}

\begin{figure*}[t]
    \centering
\includegraphics[width=0.95\linewidth]{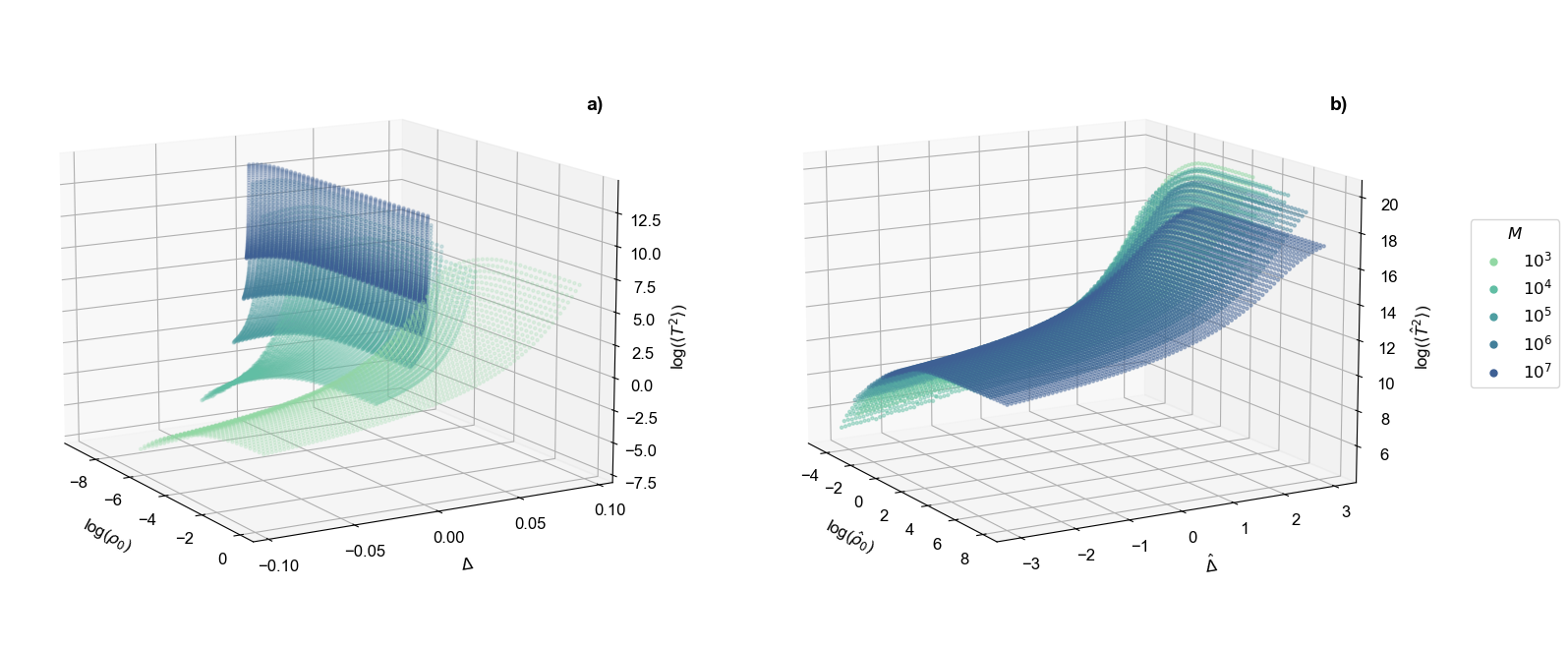} 
    \caption{\raggedright
    Second moment of exit time $\langle T^2 \rangle$ versus initial condition $\rho_0$ and deviation from criticality $\Delta$ for different values of the local carrying capacity $M$. \textbf{a)} not rescaled. \textbf{b)} rescaled according to \prettyref{eq:T2}. 
    Averages are obtained from $10^5$ numerical realizations of the MF QSS dynamics in \prettyref{eq:MFeq}.}
    \label{fig:scalingT2}
\end{figure*}

\subsection{First and second moments of exit time\label{app:1stmom}}
Consider \prettyref{eq:firstmoment} for the first moment of exit time:
\small
\begin{align}
&
\left( \Delta - \rho \right) \rho \,\partial_{\rho} T \, 
+\frac{\lambda_M \, c}{4 M} \rho
    \Big( 
    3 + \frac{\mathcal{D}}{\mathcal{D}+\lambda}
    \, - 2 \Delta \\& \qquad -\frac{2\mathcal{D}}{\mathcal{D}+\lambda}  \,\rho
    - \frac{\lambda}{\mathcal{D}+\lambda}  \rho^2 
    \Big)
    \partial_{\rho}^2 T  
    = -\frac{1}{\lambda_M \,c} \nonumber
\end{align}
\normalsize
\noindent
In order to achieve a common scaling of the LHS with the system size $M$, it is convenient to rescale $\rho \rightarrow \hat{\rho} = \rho\, \sqrt{M}$,
\small $T \rightarrow \hat{T} = \frac{\lambda_M c}{\sqrt{M}} \,T$ and 
\mbox{$\Delta \rightarrow \hat{\Delta} = \Delta \sqrt{M}$}
\begin{align}
& ( \hat{\Delta} - \hat{\rho})\hat{\rho} \,\partial_{\hat{\rho}}\hat{T} \, + \frac{1}{4}  \hat{\rho}
 \Big(  \chi  
 - 2 \,\frac{\hat{\Delta}}{\sqrt{M}}
 \, + \\ & \qquad
 - \frac{2\mathcal{D}}{\mathcal{D}+\lambda}  \frac{\hat{\rho}}{\sqrt{M}}
 - \frac{\lambda}{\mathcal{D}+\lambda}  \frac{\hat{\rho}^2}{M} 
 \Big)\partial_{\hat{\rho}}^2 \hat{T}= - 1
 \nonumber
\end{align}
\normalsize
\noindent
where we define 
$\chi \coloneqq 3 + \frac{\mathcal{D}}{\mathcal{D}+\lambda}$, which we treat as a fixed constant in the following. In the limit of large system size $M$, the terms
$- 2 \,\frac{\hat{\Delta}}{\sqrt{M}},\, 
-\frac{2\mathcal{D}}{\mathcal{D}+\lambda}  \frac{\hat{\rho}}{\sqrt{M}},
\,
 - \frac{\lambda}{\mathcal{D}+\lambda}  \frac{\hat{\rho}^2}{M} $ can be neglected,
yielding the following equation:
\begin{align}\label{eq:aclosetocrit}
 \left(\hat{\Delta} - \hat{\rho} \right)\hat{\rho}\, \partial_{\hat{\rho}} \hat{T} 
 + \frac{\chi}{4}\hat{\rho} \,\partial_{\hat{\rho}}^2 \hat{T}  = - 1 
\end{align}
which is independent of $M$. Therefore, for large $M$ the first exit time scales as:
\begin{align}
    T(\rho,\Delta,M) \propto 
    \sqrt{M}  \,\hat{T} \left( \sqrt{M} \rho, \sqrt{M} \Delta \right)
\end{align}
From the above expression we observe that the rescaled first exit time should depend only on the rescaled initial density $\hat{\rho}=\sqrt{M} \rho$ and the rescaled distance from the critical point $\hat{\Delta} = \sqrt{M} \Delta $.
The solution of \prettyref{eq:aclosetocrit} with $\hat{\rho} \in (0,\sqrt{M}]$ and boundary conditions
$\hat{T}(0) = 0 $,
$\partial_{\hat{\rho}} \hat{T}(\hat{\rho}) \rvert_{\hat{\rho}=\sqrt{M}} =  0 $ is given by the integral:
\begin{align}\label{eq:asoltocrit}
   & \hat{T}(\hat{\rho},\hat{\Delta}) = \frac{4}{\chi} \int_0^{\hat{\rho}} dx \,e^{-G(x, \hat{\Delta})}\int_x^{\sqrt{M}} dy\, \frac{e^{G(y, \hat{\Delta})}}{y}
    \nonumber \\ 
    & G(\hat{\rho}, \hat{\Delta}) = \frac{4}{\chi} \int_{\sqrt{M}}^{\hat{\rho}} dx\, (\hat{\Delta}-x) \\
    & \qquad \quad \, \, = \frac{2}{\chi } \left[2 \,\hat{\Delta} (x-\sqrt{M})+M-x^2\right] \nonumber
\end{align}
In particular, in the simple case of $\hat{\Delta}=0$ (criticality) the integral \prettyref{eq:asoltocrit} can be calculated analytically:
\begin{align}\label{eq:analyticalT}
    & \hat{T}(\hat{\rho}) =  \frac{4}{\chi}  \hat{\rho} \, _2F_2\left(\frac{1}{2},1;\frac{3}{2},\frac{3}{2};-\frac{2}{\chi}  \hat{\rho}^2\right) +\\
    & \,
     +\sqrt{\frac{\pi }{2 \chi}} \text{erfi}\left(\sqrt{\frac{2}{\chi}}  \hat{\rho} \right) \left(\text{Ei}\left(-\frac{2}{
\chi} M\right)-\text{Ei}\left(-\frac{2 }{\chi}  \hat{\rho}^2\right)\right) \nonumber
\end{align}
where $\text{erfi}(x)$ is the imaginary error function, 
$\text{Ei}(x)$ is the exponential integral function and $_p F_q (\{a_1,..a_p\};\{ b_1,...b_q\};z)$ is the generalized hypergeometric function.
\\

Consider now the second moment of extinction time ($n=2$ in \prettyref{eq:nthmoment}):
\begin{align}\label{eq:2nd}
     & \left(\Delta - \rho \right) \rho\, \partial_{\rho} T_2 
    +\frac{1}{4 M} 
    \, \rho
    \Big( 
    \chi
    - 2 \Delta \, +  \\ &
    - \frac{2\mathcal{D}}{\mathcal{D}+\lambda}  \rho
 - \frac{\lambda}{\mathcal{D}+\lambda}  \rho^2
    \Big) \,\partial_{\rho}^2 T_2  = - 2\frac{T_{1} }{\lambda_M  \, c} \nonumber
\end{align}
\noindent
where $T_1$ is the first moment, which scales according to \prettyref{eq:T1}. Rescaling $T_2  \rightarrow \hat{T}_2 = \frac{(\lambda_M c)^2}{2 M} T_2$
and neglecting terms $\sim \mathcal{O}(\frac{1}{\sqrt{M}})$,
\prettyref{eq:2nd} becomes:
\begin{align}
    \left( \hat{\Delta} - \hat{\rho} \right) \,\hat{\rho}\, \partial_{\hat{\rho}} \hat{T}_2 + \frac{\chi}{4} \hat{\rho} \,\partial_{\hat{\rho}}^2 \hat{T}_2 = - \hat{T}  
\end{align}
which shows that the second moment of exit time scales as:
\begin{align}
   T_2(\rho,\Delta,M) \propto 
   M\, \hat{T}_2\left( \sqrt{M} \rho, \sqrt{M} \Delta \right)
\end{align}

\begin{figure*}[ht!]
\centering
\includegraphics[height=0.39\linewidth]{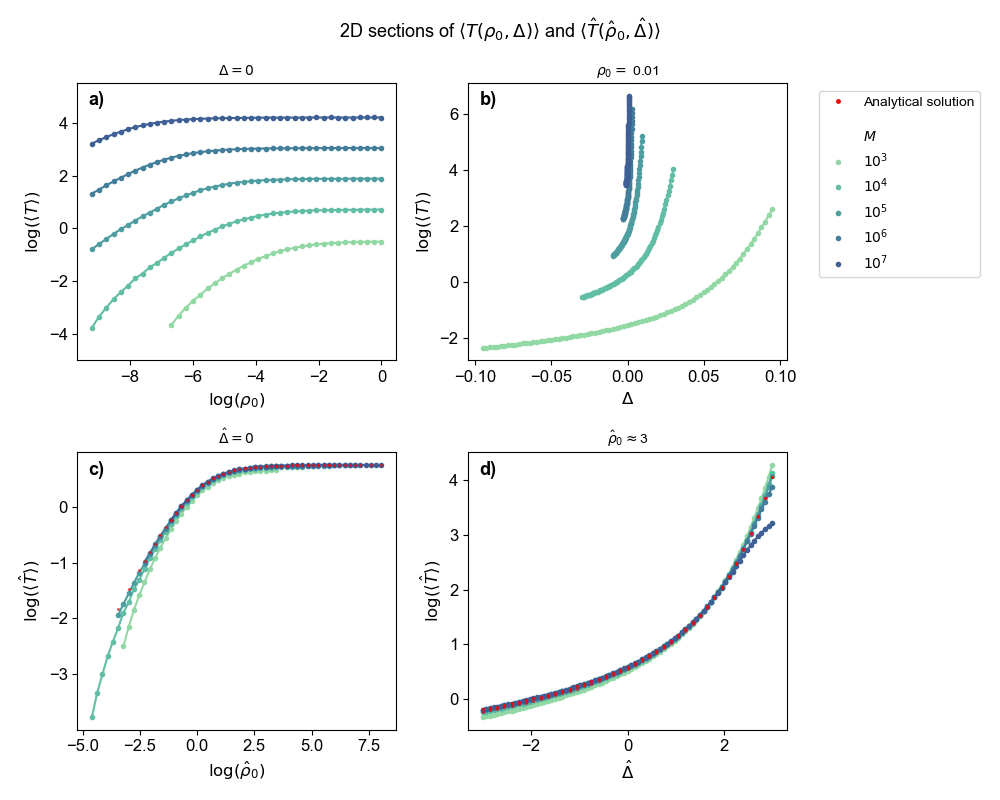}
\includegraphics[height=0.39\linewidth]{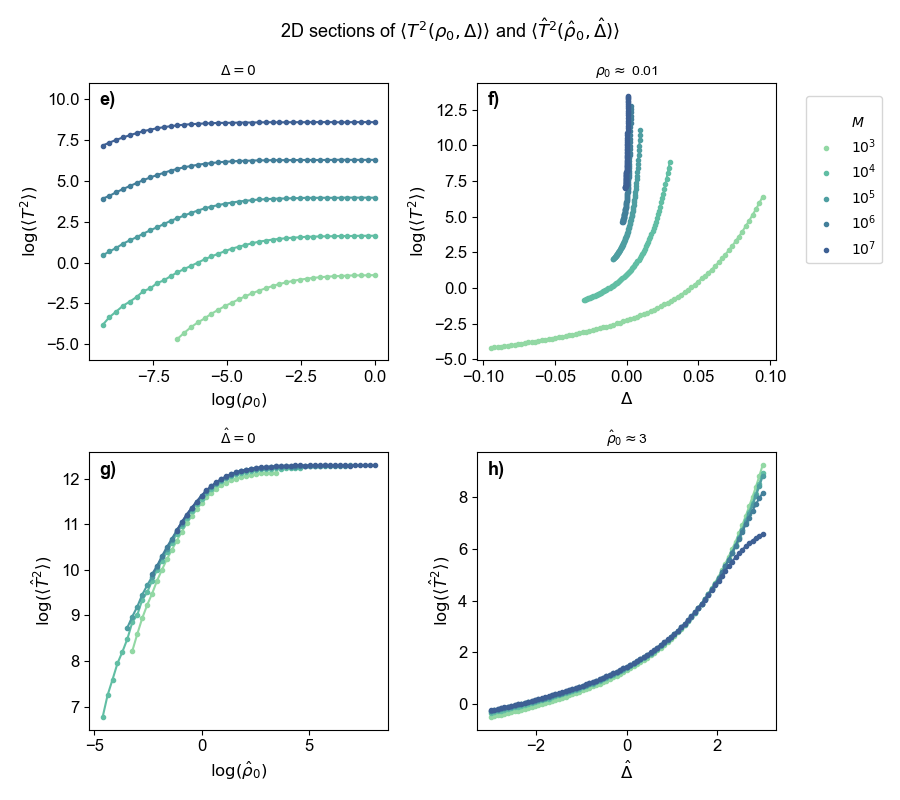}
\caption{
\raggedright
\textbf{Left:} Two-dimensional sections of \prettyref{fig:scaling}, showing the first moment of exit time  versus initial condition and deviation from criticality, for different values of the local carrying capacity $M$. \textbf{a)} Section of $\langle T\rangle$ at fixed $\Delta=0$, not rescaled. \textbf{b)} Section of $\langle T\rangle$ at fixed $\rho_0\approx0.01$, not rescaled. 
\textbf{c)} Section of $\langle \hat{T}\rangle$ at fixed $\hat{\Delta}=0$, rescaled according to \prettyref{eq:T1}.
\textbf{d)} Section of $\langle \hat{T}\rangle$ at fixed $\hat{\rho}_0\approx3$, rescaled according to \prettyref{eq:T1}.
\textbf{Right:} Two-dimensional sections of \prettyref{fig:scalingT2}, showing the second moment of exit time  versus initial condition and deviation from criticality, for different values of the local carrying capacity $M$.
\textbf{e)} Section of $\langle T^2\rangle$ at fixed $\Delta=0$, not rescaled. \textbf{e)} Section of $\langle T^2\rangle$ at fixed $\rho_0\approx0.01$, not rescaled. 
\textbf{f)} Section of $\langle \hat{T}^2\rangle$ at fixed $\hat{\Delta}=0$, rescaled according to \prettyref{eq:T2}.
\textbf{g)} Section of $\langle \hat{T}^2\rangle$ at fixed $\hat{\rho}_0\approx3$, rescaled according to \prettyref{eq:T2}.
}
\label{fig:slices}
\end{figure*}

\indent
The analytical solution is given by:
\begin{align}\label{eq:aT2integral}
    \hat{T}_2(\hat{\rho}, \hat{\Delta}) = \frac{4}{\chi}\int_0^{\hat{\rho}} dx \,e^{-G(x,\hat{\Delta})} \int_x^{\sqrt{M}} dy \, \frac{\hat{T}(y,\hat{\Delta})}{y} e^{G(y,\hat{\Delta})}
\end{align}
\indent
with $G(\hat{\rho}, \hat{\Delta})$ and $\hat{T}(\hat{\rho}, \hat{\Delta})$ defined in \prettyref{eq:asoltocrit}.
 As done for the mean extinction time, we numerically calculate the second moment of extinction time by simulating $10^5$ realizations of the mean-field QSS dynamics given in \prettyref{eq:MFeq}, using the Euler-Maruyama algorithm 
\cite{Kloeden1992}.
\mbox{By applying} the proper rescaling, we obtain again a good collapse of the second moment of extinction time as a function of the distance from the critical point $\Delta$, the initial condition $\rho$ and the system size $M$. The collapse is shown in \prettyref{fig:scalingT2}, and clarified through two-dimensional sections in the right panel of \prettyref{fig:slices}.

\begin{figure}[h!]
\centering
\includegraphics[width=\linewidth]{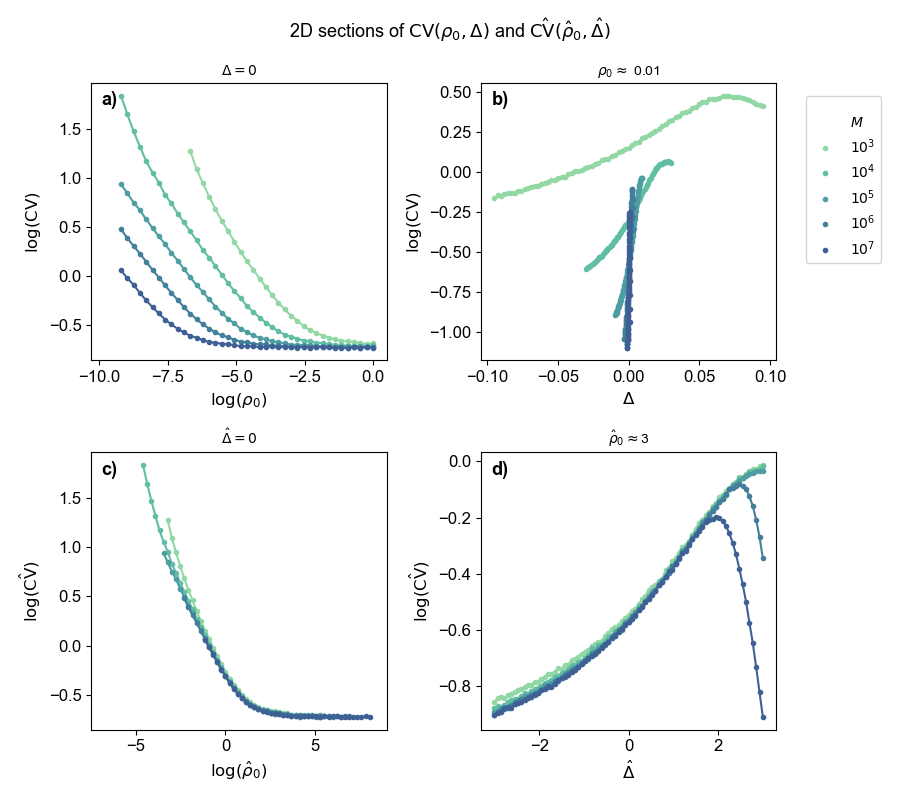}
\caption{
\raggedright
\textbf{Left:} Two-dimensional sections of the top panel of \prettyref{fig:eco}, showing the coefficient of variation (CV) as a function of the initial condition $\rho_0$ and deviation from criticality $\Delta$, for different values of the local carrying capacity $M$.  
\textbf{a)} Section of $\langle \mathrm{CV} \rangle$ at fixed $\Delta = 0$, not rescaled.  
\textbf{b)} Section of $\langle \mathrm{CV} \rangle$ at fixed $\rho_0 \approx 0.01$, not rescaled.  
\textbf{c)} Section of $\langle \hat{\mathrm{CV}} \rangle$ at fixed $\hat{\Delta} = 0$, rescaled according to \prettyref{eq:T1} and \prettyref{eq:T2}.  
\textbf{d)} Section of $\langle \hat{\mathrm{CV}} \rangle$ at fixed $\hat{\rho}_0 \approx 3$, rescaled according to \prettyref{eq:T1} and \prettyref{eq:T2}.
}
\label{fig:CVslices}
\end{figure}

\subsection{Survival probability\label{app:survprob}}
 Starting from 
 \prettyref{eq:survprob}, we apply the rescaling procedure $
\rho \rightarrow \hat{\rho}=  \rho\, M^{\gamma},
\, t \rightarrow \hat{t}=t \lambda_M c M^\nu 
$, which yields:
\small
\begin{align}
   & M^\nu\,\partial_{\hat{t}} S = M^{-\gamma} (\Delta M^\gamma- \hat{\rho})  \hat{\rho} \,\partial_{\hat{\rho}} S \, + \\ &
    +\frac{1}{4}M^{\gamma-1}
    \hat{\rho}
    \left(
    \chi - 2\Delta
    - \frac{2\mathcal{D}}{\mathcal{D}+\lambda}  \frac{\hat{\rho}}{M^\gamma}
 - \frac{\lambda}{\mathcal{D}+\lambda}  \frac{\hat{\rho}^2}{M^{2\gamma}} 
    \right)
    \partial{\hat{\rho}}^2 S \nonumber 
\end{align}
\normalsize
It is then convenient to choose $\nu,\, \gamma$ such that $\nu = - \gamma = \gamma-1 \,\rightarrow \,\, \gamma=\frac{1}{2}, \, \nu = -\frac{1}{2}$.
\begin{align}
    & \partial_{\hat{t}} S =  \,
    (\hat{\Delta} - \hat{\rho})  \hat{\rho} \,\partial_{\hat{\rho}} S\,+  \\
    & 
    + \frac{1}{4} 
 \left( 
 \chi  
 - 2 \,\frac{\hat{\Delta}}{\sqrt{M}}  
 - \frac{2\mathcal{D}}{\mathcal{D}+\lambda}  \frac{\hat{\rho}}{\sqrt{M}}
 - \frac{\lambda}{\mathcal{D}+\lambda}  \frac{\hat{\rho}^2}{M} 
 \right)\hat{\rho} \,
    \partial{\hat{\rho}}^2 S
    \nonumber
\end{align}
where $\hat{\Delta}=\Delta \sqrt{M}$.
Neglecting  $\mathcal{O}(\frac{1}{\sqrt{M}})$ terms:
\begin{align}
    \partial_{\hat{t}} S = 
    (\hat{\Delta} - \hat{\rho})  \hat{\rho} \,\partial_{\hat{\rho}} S
    +\frac{\chi}{4} \hat{\rho} \,\partial{\hat{\rho}}^2 S
\end{align}
We deduce the following scaling behavior of the survival probability:
\begin{align}
   S(t,\rho,\Delta,M) = S\left( \frac{t}{\sqrt{M}}, \sqrt{M}\rho, \sqrt{M}\Delta \right)
\end{align}
which is consistent with the scaling of $S$ determined through the previous approach.

\FloatBarrier
\bibliography{biblio}

\end{document}